\documentclass[a4paper,11pt]{article}
\pdfoutput=1 

\usepackage{jcappub} 

\usepackage{ulem}
\usepackage{amsmath}
\usepackage{amsfonts,color}
\usepackage{amssymb,float}
\usepackage[utf8]{inputenc}
\usepackage{enumerate}
\usepackage{mathrsfs}

\usepackage{color}

\usepackage{hyperref}
\usepackage{enumitem}

\newcommand{\Lag}{{\mathcal L}}
\newcommand{\mS}{{\mathcal S}}
\newcommand{\mH}{{\mathcal H}}
\newcommand{\mK}{{\mathcal K}}
\newcommand{\mP}{{\mathcal P}}
\newcommand{\mM}{{\mathcal M}}
\newcommand{\mL}{{\mathcal L}}
\newcommand{\mA}{{\mathcal A}}

\newcommand{\mJ}{{\mathcal J}}
\newcommand{\mj}{\rho}

\newcommand{\dualH}{\tilde{H}}
\newcommand{\dualP}{\tilde{P}}

\newcommand{\phib}{\bar{\phi}}

\newcommand{\Bb}{\bar{B}}
\newcommand{\Yb}{\bar{Y}}

\newcommand{\cs}{c_{\text s}}

\newcommand{\be}{\begin{equation}}
\newcommand{\ee}{\end{equation}}

\newcommand{\bea}{\begin{eqnarray}}
\newcommand{\eea}{\end{eqnarray}}

\newcommand{\nn}{\nonumber \\}

\newcommand{\dd}{{\rm d}}

\begin{document}

\title{Some disquisitions on cosmological 2-form dualities}

\author[a]{Katsuki Aoki,}
\author[b]{Jose Beltr\'an Jim\'enez,}
\author[b]{David Figueruelo}

\affiliation[a]{Center for Gravitational Physics and Quantum Information, Yukawa Institute for Theoretical Physics, Kyoto University, 606-8502, Kyoto, Japan}
\affiliation[b]{Departamento de F\'isica Fundamental and IUFFyM, Universidad de Salamanca, E-37008 Salamanca, Spain}

\emailAdd{katsuki.aoki@yukawa.kyoto-u.ac.jp}
\emailAdd{jose.beltran@usal.es}
\emailAdd{davidfiguer@usal.es}

{\baselineskip0pt
\rightline{\baselineskip16pt\rm\vbox to-20pt{
           \hbox{YITP-22-158}
\vss}}%
}

\abstract{In this work we study different aspect of self-interacting 2-form fields with special emphasis in their cosmological applications. We provide the explicit construction of how massless 2-forms are compatible with the cosmological principle without resorting to the dual scalar field formulation. In terms of the 2-form, the residual Euclidean group is non-trivially realised by means of a combination of external spatial translations and internal gauge transformations. After presenting the general discussion of the dualities in cosmological scenarios, we analyse particular examples for some singular models and discuss in some detail the dual descriptions of the DBI, the cuscuton and the ghost condensate as well as the role of the duality in the effective field theories of cosmological perturbations. We then proceed to analysing scenarios with several self-interacting massless 2-forms and we show that they naturally provide the dual description of a solid. We then show how the perfect fluid and superfluids can be obtained by taking the appropriate limits in the dual formulations. We finally consider the case of massive 2-forms and their duals and briefly discuss their potential signatures in gravitational waves astronomy.}
\date{\today}


\maketitle
\newpage
\section{Introduction}

The cosmological principle dictates that our universe is spatially homogeneous and isotropic on sufficiently large scales and represents a major pillar of the standard cosmological model.\footnote{The validity of the cosmological principle has been challenged in e.g.\cite{Secrest:2022uvx} (see however \cite{Dalang:2021ruy,Guandalin:2022tyl}) from the observation of a dipole in the distribution of radio galaxies and quasars that differ from the CMB one with a statistical significance of about $5\sigma$. A result along these lines has also been obtained in \cite{Sorrenti:2022zat} by using the Pantheon+ supernovae catalogue. Similar conclusions had also been reached in \cite{Atrio-Barandela:2014nda} where the existence of a large scale dark flow was claimed. Although these are all intriguing observational findings, more robust (better control on the systematics) and statistically significant results are needed before abandoning one of the pillars of the standard model. Thus, we commit to the cosmological principle in this work.} This principle is of paramount importance for developing theoretical cosmological models since it requires compliance with the required symmetries. In other words, the proposed models should allow for background solutions realising the symmetries imposed by the cosmological principle, i.e., they need to allow for solutions invariant under a residual 3-dimensional Euclidean group $ISO(3)$. This requirement explains why cosmological models based on scalar fields are by far the most extensively explored class of theories with applications in cosmology, since it is trivial to realise the cosmological principle by simply taking a homogeneous scalar field profile. Though this is arguably the simplest realisation, it is not the only one and realisations of the cosmological principle where the residual $ISO(3)$ symmetry is achieved by means of a combination of external and internal symmetries also exist. Yang-Mills theories possessing an internal $SU(2)$ symmetry admit vacuum configurations that break isotropy and the internal group simultaneously, but they leave a linear combination of the two unbroken, thus preserving a diagonal rotational group. These configurations were explored in cosmological scenarios as early as \cite{Cervero:1978db} and have found numerous applications (see e.g. \cite{Galtsov:1991un,Darian:1996mb,Maleknejad:2011jw,Maleknejad:2012fw} among others). Spin-1 fields that break the gauge symmetry but retain a global $SO(3)$ invariance also allow for the discussed realisation of the cosmological principle as in the dark energy model introduced in \cite{Armendariz-Picon:2004say}. All these scenarios realise homogeneity in a trivial way because the background configurations only depend on time. It is however also possible to construct models with a non-trivial realisation of homogeneity. Among these models, we can mention {\it solid} cosmologies \cite{Endlich:2012pz} (see also the earlier developments \cite{Bucher:1998mh,Gruzinov:2004ty}) that can be described in terms of three scalar fields with an internal $ISO(3)$ symmetry and the homogeneity is achieved as a combination of spatial and internal translations. Another class of models with non-trivially realised homogeneity is provided by the gaugid cosmologies \cite{Piazza:2017bsd} that rely on massless spin-1 fields with a global $SO(3)$ symmetry so homogeneity combines spatial translations and the three Abelian gauge symmetries. For an exhaustive classification of possible combinations of Lorentz and internal generators relevant for non-trivial realisations of the cosmological symmetries see \cite{Nicolis:2015sra}. So far, we have mentioned models with continuous internal symmetries, but it is remarkable that even internal discrete symmetries can make the case for homogeneous and isotropic cosmologies \cite{Kang:2015uha}. The study of models with these different (and inequivalent) realisations of the cosmological principle do not represent unnecessary complications for describing our universe, but each of them gives rise to crucially different phenomenologies with clear signatures. Furthermore, they also permit us to advance in our theoretical understanding of the underlying structure of the cosmological models.

Another interesting possibility that allows to advance in our understanding of the cosmological models is the existence of dualities between different realisations of the cosmological principle. In fact, establishing duality relations between seemingly inequivalent realisations of the cosmological symmetries explained above might help understanding the underlying structure of such configurations. For instance, the symmetries that appear non-trivially realised in a given model, turn out to be trivially realised in the dual description. Furthermore, the dual realisation of the symmetries might unveil additional hidden symmetries in the original realisation. The aim of this work is exploring some existing dualities for the realisations of the cosmological principle by resorting to the known dualities that exist among different order $p$-form fields. It is well-known that a massless 2-form (Kalb-Ramond) field can be dualised to a shift symmetric theory for a scalar field \cite{Cremmer:1973mg}. Thus, it is customary to go to the dual formulation in terms of a scalar field to study cosmologies with massless 2-form fields. In this dual formulation, it is obvious that the background solution is provided by a homogeneous scalar field. This duality has been exploited to construct cosmological solutions in e.g. \cite{Stein-Schabes:1986owe,Copeland:1994km,Elizalde:2018rmz}. It is interesting however to notice that the dual formulation with the Kalb-Ramond field requires the use of its gauge symmetry to properly realise the residual $ISO(3)$ symmetry. This is a first example of a relation between different realisations of the cosmological principle that could seem unrelated a priori. We will work out in detail how the cosmological solutions are described in terms of the 2-form and, in particular, how the symmetries are realised. We will explicitly show that the cosmologies of a self-interacting Kalb-Ramond field are completely equivalent to those of a shift-symmetric $K$-essence theory for a scalar field and work out the relations between the relevant cosmological quantities (equation of state parameter, sound speed, etc.) in both descriptions. 

We will then proceed to a more interesting scenario involving several self-interacting Kalb-Ramond fields. Providing these fields with an internal $SO(3)$ symmetry, in addition to the three gauge symmetries of each massless 2-form, allows for a class of cosmologies with a non-trivially realised cosmological principle that is to some extent similar to those constructed with Yang-Mills fields. Interestingly, these configurations are nothing but the dual formulation of the solid cosmologies. We will see how the configuration in terms of the two form seems more natural from the point of view of the symmetries because homogeneity is trivially realised, unlike in the formulation with scalar fields where the internal translations are necessary to preserve homogeneity. In other words, while the solid cosmologies with three scalar fields require inhomogeneous configurations for the ground state, the dual formulation in terms of 2-forms is homogeneous.

After exploring at length the dualities existing between shift-symmetric scalar field theories and massless Kalb-Ramond fields, we will make a brief incursion into the massive 2-form theories. For these massive theories, as usual, the phase space is enhanced  due to the breaking of the gauge symmetry and the dual description necessitates a massive spin 1 field \cite{Curtright:1980yj}. The absence of internal symmetries makes the dual formulations more rigid since we can no longer play with internal symmetries. As a matter of fact, constructing cosmological configurations is more difficult, precisely due to the absence of additional symmetries and, for the simplest models, they all boil down to pure trivial solutions with, at most, a cosmological constant. Introducing several massive 2-form fields brings back the possibility of playing with internal symmetries and we will see how these internal symmetries permit again non-trivial realisations of the cosmological principle that are dual to the configurations that can be constructed with massive spin-1 fields. These scenarios bring about novel phenomenological effects due to the presence of an additional helicity-2 tensor as we will show.

The paper is organised as follows. In Sec.~\ref{sec:KRcosmology}, we sketch out the basic equations, the symmetry, and the cosmological configuration of a self-interacting massless 2-form. We then elaborate on the duality between a massless 2-form field and a massless scalar field in Sec.~\ref{sec:dualities} and provide several examples in Sec.~\ref{sec:examples}. We revisit the cosmology in the massless 2-form in Sec.~\ref{sec:cosmologicalduality} with emphasis in the duality relation to the cosmology in the scalar field. Sec.~\ref{sec:multiKR} is devoted to a multi-field extension and the duality between the massive 2-form and the massive vector field is discussed in Sec.~\ref{sec:massive2form}. We conclude in Sec.~\ref{sec:discussion} with summary and discussions.

\section{Kalb-Ramond cosmologies}
\label{sec:KRcosmology}
We will commence our disquisitions on 2-forms by reviewing some basic properties of Kalb-Ramond fields. We will also use this section to discuss in detail how the Kalb-Ramond field can provide a non-trivial realisation of the cosmological principle by making use of its gauge symmetry. For simplicity, we focus on the flat spacetime in this section but it can capture the essential properties and the inclusion of gravity is straightforward.

\subsection{Generalities}
Let us consider a theory for a self-interacting Kalb-Ramond field $B_{\mu\nu}$ described by the action
\be
\mS=\int\dd^4x
F(Y)
\label{eq:actionF}
\ee
where we have defined $Y\equiv-\frac{1}{12} H_{\mu\nu\rho} H^{\mu\nu\rho}$ with $H_{\mu\nu\rho}=3 \partial_{[\mu} B_{\nu\rho]}$ the corresponding gauge-invariant field strength. This is a good effective field theory because quantum corrections will only enter with higher derivatives as $\partial^nY$ with $n\geq1$.
This theory possess the usual gauge symmetry of a massless Kalb-Ramond field, namely:
\be
B_{\mu\nu}\rightarrow B_{\mu\nu}+2\partial_{[\mu}\theta_{\nu]}
\label{eq:gaugesym1}
\ee
for an arbitrary $\theta_\nu$. Let us notice that the gauge transformation itself has a gauge symmetry so that $\theta_\mu$ and $\theta_\mu+\partial_\mu \theta$ describe the same gauge transformation. Since the Kalb-Ramond field is an antisymmetric field, it is convenient to parameterise it with two 3-vectors as
\bea
B_i\equiv B_{0i},\quad
C^i\equiv\frac12\epsilon^{ijk} B_{jk}\,,
\eea
that can be called the electric and magnetic components of the Kalb-Ramond field respectively due to their behaviour under parity transformations, in analogy to the electromagnetic field. The corresponding field strength can be written as
\be
H_{0ij}=\epsilon_{ijk}\dot{C}^k-2\partial_{[i}B_{j]}\,,\quad H_{ijk}=\nabla\cdot\vec{C}\,\epsilon_{ijk}.
\ee
From this expression it is already apparent that the longitudinal part of the electric piece of the Kalb-Ramond field $\vec{B}$ does not contribute to the field strength so it does not play any physical role. As we will see in the following, this makes that, out of the six components of the two form field, only five of them are actually relevant. In other words, the longitudinal part of $\vec{B}$ is entirely associated to the longitudinal component of the gauge transformation so this sector is trivial, unlike the transverse modes that realise the symmetry in a non-trivial way. The vectors $\vec{B}$ and $\vec{C}$ transform under the gauge symmetry as follows:
\begin{align}
\vec{B}&\rightarrow\vec{B}+\dot{\vec{\theta}}-\nabla\theta_0\,,\\
\vec{C}&\rightarrow\vec{C}+\nabla\times \vec{\theta}\,,
\end{align}
where we see that we can change $\vec{\theta}\rightarrow \vec{\theta}+\nabla\theta$ and $\theta_0\rightarrow \theta_0+\dot{\theta}$ without affecting the transformation of the fields. We can use this freedom to set e.g. $\theta_0=0$ and describe the gauge freedom with the three parameters of $\vec{\theta}$. Furthermore, we can see that $\vec{C}$ only transforms with the transverse part of $\theta$ and its longitudinal component is gauge invariant. This is the single dynamical degree of freedom of the theory. To see this more explicitly, it is convenient to decompose the fields into transverse and longitudinal components
\be
\vec{B} =\vec{B}^T + \nabla B
\,, \quad
 \vec{C} =  \vec{C}^T + \nabla C
 \,,
\ee
with $\nabla \cdot \vec{B}^T = \nabla \cdot \vec{C}^T =0$. The transformation of these components under the gauge symmetry is given by
\begin{align}
\vec{B}^T &\to \vec{B}^T + \dot{\theta}^T 
\,, \\
B & \to  B + \dot{\theta} -  \theta_0
\,, \\
\vec{C}^T &\to\vec{C}^T + \nabla \times \vec{\theta}^T
\,, \label{CTgaugeM} \\
C & \to  C\,,
\end{align}
where we have also decomposed the gauge parameter as $\theta_\mu=\big(\theta_0,\vec{\theta}^T+\nabla\theta\big)$. Since, as we have shown above, the longitudinal mode $B$ does not appear in the field equations, it is a superfluous component of the Kalb-Ramond field and only five out of its six components actually contribute to the field equations. Furthermore, this superfluous mode also carries the gauge freedom of the longitudinal mode of the gauge parameter so that we only have a two-dimensional gauge symmetry spanned by $\vec{\theta^T}$.\footnote{This is consistent with the fact that the equations \eqref{Constrainteq} below provide two constraint equations that are associated to the gauge symmetry generated by $\theta^T$.} Thus, the counting of degrees of freedom is as follows: 5 components of the Kalb-Ramond field that contribute to the equations, minus the two constraints, minus the two gauge freedoms, accounting for a total of one dynamical degree of freedom, as it should.

The field equations of the Kalb-Ramond field read
\be
\partial_\mu\Big (F_YH^{\mu\alpha\beta}\Big)=J^{\alpha\beta}\,,
\ee
where we have included a source $J^{\mu\nu}=J^{[\mu\nu]}$ that is subject to the conservation law
\be
\partial_\alpha J^{\alpha\beta}=0,
\ee
consistent with the gauge symmetry of the Kalb-Ramond field. This antisymmetric current can be analogously decomposed as
\bea
j_i\equiv J_{0i},\quad
\mJ_i\equiv\frac12\epsilon^{ijk} J_{jk}\,,
\eea
so the conservation law reads
\be
\frac{\partial \vec{j}}{\partial t}+\nabla\times\vec{\mJ}=0,\,\quad \nabla\cdot\vec{j}=0\,.
\ee
The latter of the above equations implies that there must be some $\vec{\mj}$ such that $\vec{j}=\nabla\times\vec{\mj}$ on-shell so  from the first equation we further find that $\dot{\vec{\mj}}+\vec{\mJ}=\nabla\mj$, for some scalar function $\mj$. The field equations can also be written in terms of the electric and magnetic components of the Kalb-Ramond field as follows:
\begin{align}
\nabla\times\left[F'\Big(\dot{\vec{C}}-\nabla\times\vec{B}\Big)\right]=&\vec{j}\,,\label{Constrainteq}\\
-\partial_0\left[F'\Big(\dot{\vec{C}}-\nabla\times\vec{B}\Big)\right]+\nabla\Big(F'\nabla\cdot\vec{C}\Big)=&\vec{\mJ}\,.
\end{align}
We can use the obtained constraints for the current to rewrite these equations as
\begin{align}
\nabla\times\left[F'\Big(\dot{\vec{C}}-\nabla\times\vec{B}\Big)-\vec{\mj}\,\right]=&0\,,\label{Constrainteq2}\\
\partial_0\left[\vec{\mj}-F'\Big(\dot{\vec{C}}-\nabla\times\vec{B}\Big)\right]+\nabla\Big(F'\nabla\cdot\vec{C}-\mj\Big)=&0\,.
\end{align}
These equations serve as the starting point for some formal analysis and phenomenological studies,\footnote{For instance, these equations for the theory without self-interactions of the Kalb-Ramond field (i.e. $F_Y=1$) have been used in \cite{Matsuo:2021xas} to show a formal resemblance with the Euler equations, thus providing a description of a perfect fluid with a Kalb-Ramond field.} although we will not develop it any further here. Instead, we will now proceed to show how the cosmological principle can be non-trivially realised by using the gauge symmetry of the Kalb-Ramond field.

\subsection{Cosmological configurations: Non-trivially realising the cosmological principle}

In order to have a homogeneous and isotropic cosmological solution, we will consider the following field configuration
\be
B_{ij}=\frac13 B\epsilon_{ijk} x^k\;,\quad B_{0i}=0\,,
\label{eq:fieldconf}
\ee
with $B$ some constant and $\epsilon_{ijk}$ the $SO(3)$-invariant completely antisymmetric tensor. By a suitable global rescaling of the spatial coordinates we can set $B=1$, but we will keep it explicitly for convenience to keep track of the background effects. At first sight, this field configuration seems to break both homogeneity and isotropy. However, it does preserve those symmetries thanks to the gauge symmetry of the theory. If we simultaneously perform a spatial translation plus a gauge transformation we obtain
\be
B_{ij}\rightarrow B_{ij+}\frac13 B \epsilon_{ijk}x_0^k+2\partial_{[i}\theta_{j]}
\ee
or, in terms of $\vec{C}$,
\be
\vec{C}\rightarrow\vec{C}+\frac{1}{3} B\vec{x}_0+\nabla\times\vec{\theta}\,.
\ee
Since we can express $\vec{x}_0=\frac12\nabla\times(\vec{x}_0\times\vec{x})$ we see that if we choose $\vec{\theta}=\frac16 B\vec{x}\times\vec{x}_0$, the above variation vanishes and we can realise homogeneity by means of a combination of spatial translations and gauge transformations.  Notice that this corresponds to a {\it large gauge transformation}, i.e., it does not vanish at infinity. This mechanism is similar to the magnetic gaugid considered in \cite{Piazza:2017bsd}. One can proceed similarly to check that a combination of spatial rotations and gauge transformations remains unbroken as to realise isotropy. For that, it is simpler to use the Euclidean 3-dimensional Hodge dual of the two form $C^i=\frac12\epsilon^{ijk}B_{jk}=\frac B3 x^i$. Under a spatial rotation described by $R^i{}_j\simeq\delta^i{}_j+\omega^i{}_j$, the background configuration $C^i=\frac B3x^i$ changes as
\be
C^i(x)\rightarrow R^i{}_jC^j(R\cdot x)\simeq C^i+2\omega^i{}_jC^j\,.
\ee
The corresponding transformation for $B_{ij}$ is\footnote{The transformation of $B_{ij}$ can be obtained directly from 
\be
B_{ij}\rightarrow R_i{}^m R_j{}^n B_{mn}(R\cdot x)\,,
\ee
and using the relation $R^i{}_mR^j{}_n\epsilon_{ijk}=\epsilon_{mn\ell}R_p{}^\ell$ obtained from $\det R=1$ that gives the corresponding identity for the generators $\epsilon_{inp}\omega_m{}^i+\epsilon_{mjp}\omega_n{}^j=\epsilon_{mn\ell}\omega^\ell{}_p$. The final result is of course the same but it is substantially simpler to work with $C^i$ directly.}
\be
\delta B_{ij}=\epsilon_{ijk}\delta C^k=\frac23 B\epsilon_{ijk}\omega^k{}_\ell x^\ell\,.
\ee

We then have that the action of an infinitesimal spatial rotation and a gauge transformation is given by
\be
\vec{C}\rightarrow\vec{C}-\frac23 B\vec{\omega}\times\vec{x}+\nabla\times\vec{\theta}\,,
\ee
where $\vec{\omega}$ is the vector associated to the rotation so that $\omega^i{}_j x^j=\epsilon^i{}_{jk}\omega^kx^j=-(\vec{\omega}\times\vec{x})^i$. If we now use that $\vec{x}=\frac12\nabla\vec{x}^2$ and $\nabla\times(\vec{x}^{\,2}\vec{\omega})=-\vec{\omega}\times\nabla\vec{x}^2$, we finally obtain
\be
\vec{C}\rightarrow\vec{C}+\nabla\times\left(\frac13 B\vec{x}^2\vec{\omega}+\vec{\theta}\right)
\ee
so that rotations are realised upon choosing the gauge transformation parameter as $\vec{\theta}=-\frac13 B\vec{x}^2\vec{\omega}$. This gauge transformation is even larger than the one necessary to realise translations as it grows one power faster. Thus, the background configuration is invariant under the diagonal $ISO_{\rm d}(3)$ group realised by the following combination of coordinate and gauge transformations:
\bea
\delta x^i&=&\omega^i{}_jx^j+x_0^i\,,\\
\theta_i&=&\frac{B}{6}\epsilon_{ijk}\big(x^jx_0^k-\vec{x}^2\omega^{jk}\big),\quad\theta_0=0\,.
\eea
It is important to notice that $B_{0i}$ does not change under the residual $ISO_{\rm d}(3)$ since
\be
B_{0i}\rightarrow B_{0i}+\partial_0\theta_i-\partial_i\theta_0=B_{0i}\,.
\ee

Perhaps a more direct manner to corroborate the homogeneity and isotropy of the cosmological solutions generated by \eqref{eq:fieldconf} is to notice that the physical quantity is not the 2-form, but its field strength, which is given by
\be
H_{ijk}=B\epsilon_{ijk}
\label{Hconfig}
\ee
which obviously realises the residual $ISO_{\rm d}(3)$ symmetry. The energy-momentum tensor is given by
\be
T_{\mu\nu}=\frac12 F_Y H_{\mu\alpha\beta}H_\nu{}^{\alpha\beta}+F g_{\mu\nu}
\ee
which, in the field configuration \eqref{eq:fieldconf}, is homogeneous and isotropic with components $T^\mu{}_\nu=\text{diag}(-\rho,p,p,p)$ where the energy density and pressure are given by
\begin{align}
\rho=& -F\,,\\
p=&F\left(1-2\frac{\partial\log F}{\partial \log Y}\right)\,.
\end{align}
We then see that the condition to have inflationary solutions is
\be
\frac{\partial\log F}{\partial \log Y}\ll1\,.
\ee
We will be more precise on the slow roll conditions below, but let us first explore the theory on a Minkowski background. Due to the decoupling theorem of gravitational degrees of freedom at small scales, this will be the relevant regime for modes well below the Hubble horizon. In our background field configuration \eqref{eq:fieldconf}, we have that $Y=\Yb$ is constant and the background field equation is trivially satisfied, i.e., the constant $\Yb$ is only determined by boundary conditions. Around this background, we will consider perturbations of the 2-form field
\be
B^{(1)}_{0i}= \delta B_i,\quad B_{ij}^{(1)}=\epsilon_{ijk} \delta C^k.
\ee
It is convenient to decompose the perturbations into the transverse modes and the longitudinal modes
\be
\delta \vec{B}_i = \delta \vec{B}^T + \nabla \delta B
\,, \quad
\delta \vec{C}_i = \delta \vec{C}^T + \nabla \delta C
\,,
\ee
with $\nabla \cdot \delta \vec{B}^T = \nabla \cdot \delta \vec{C}^T =0$. The gauge transformations are given by
\bea
\delta \vec{B}^T &\to& \delta \vec{B}^T + \delta \dot{\theta}^T 
\,, \\
\delta B & \to & \delta B + \delta \dot{\theta} - \delta \theta_0
\,, \\
\delta \vec{C}^T &\to& \delta \vec{C}^T + \nabla \times \delta \vec{\theta}^T
\,, \label{CTgauge} \\
\delta C & \to & \delta C
\,,
\eea
where the spatial part of the gauge parameter has also been decomposed as $\delta \vec{\theta}=\delta \vec{\theta}^T + \nabla \delta \theta$, as done before for the background. The transformation rules show that $\delta C$ and the combination $\delta \dot{\vec{C}}^T-\nabla \times \delta \vec{B}^T$ are the gauge-invariant quantities.

The quadratic action in Minkowski spacetime for perturbations around this background is given by 
\bea
\mS=\frac{1}{2}\int\dd^4x\left[F_Y\Big(\nabla \delta \dot{C} + \delta\dot{\vec{C}}^T-\nabla\times\delta\vec{B}^T \Big)^2-\Big(F_Y+2YF_{YY}\Big)(\nabla\cdot \nabla \delta C)^2\right] \,,
\label{2-formqaction1}
\eea 
which is written in terms of the gauge-invariant quantities as it should be. We now fix the gauge so that $\delta\vec{C}^T=0$ by choosing the gauge parameter $\delta \vec{\theta}^T$.\footnote{Although one may choose another gauge $\delta \vec{B}^T=0$, there is a freedom associated with the initial condition of $\delta \vec{\theta}^T$; thus, this gauge choice does not completely fix the gauge. On the other hand, $\delta\vec{C}^T=0$ provides a complete gauge-fixing under an appropriate boundary condition. Alternatively, one can use the gauge-invariant variable $\delta \vec{G}^T$ defined by $\nabla\times \delta \vec{G}^T=-\delta \dot{\vec{C}}^T+\nabla \times \delta \vec{B}^T$ without gauge-fixing.}
We see the non-dynamical character of $\delta\vec{B}^T$ and the constraint equation generated by $\delta\vec{B}$ is written as
\be
\nabla\times\nabla\times\delta\vec{B}^T=0 \quad \implies \quad \delta\vec{B}^T=0
\,.
\ee
If we plug this solution back in the action, expressed in Fourier space for simplicity, we obtain
\be
\mS=\frac{1}{2}\int\dd t\dd^3kk^2F_Y\left[\delta\dot{C}_k^2-\Big(1+\frac{2YF_{YY}}{F_Y}\Big)k^2\delta C_k^2 \right]\,.
\label{2-formqaction2}
\ee
This expression clearly shows that the theory contains one scalar mode, in accordance to the fact that a 2-form field is dual to a scalar field. We will discuss and exploit this duality in more detail below. The obtained quadratic action allows to easily identify the conditions for the absence of ghosts and Laplacian instabilities, which are
\be
F_Y>0\,,\quad\cs^2=1+\frac{2YF_{YY}}{F_Y}>0\,.
\label{eq:stabilityMinkowski}
\ee
While the first condition is easily satisfied for interesting cosmologies, we will see later that the absence of Laplacian instabilities is at odds with the existence of slow-roll inflationary solutions. Before that, let us explore the duality relations of the Kalb-Ramond field.

\section{Dualities}
\label{sec:dualities}

In this section we will explicitly construct the duality relation between the self-interacting Kalb-Ramond field introduced in the preceding section and the shift-symmetric $K$-essence theories. We will discuss the weak/strong coupling relation for the dual descriptions and how it can be understood from the perspective of fluids.

\subsection{Dualisation}
\label{sec:dualisation}
Let us consider a theory for a Kalb-Ramond field in the first order formalism
\be
\mS=-\frac12\int\dd^4x\sqrt{-g}\left[\Pi^{\mu\nu\rho} \partial_{[\mu} B_{\nu\rho]}-\frac16 \Pi^2\right] \,.
\ee
Upon variation w.r.t. the conjugate momentum $\Pi^{\mu\nu\rho}$ we obtain the equation 
\be
\Pi_{\mu\nu\rho}=3\partial_{[\mu} B_{\nu\rho]}\equiv H_ {\mu\nu\rho}
\ee
that, when plugged back into the action, recovers the second order formulation. On the other hand, the equation for the 2-form gives
\be
\nabla_\mu \Pi^{\mu\nu\rho}=0
\ee
that is solved by
\be
\Pi^{\mu\nu\rho}=\varepsilon^{\mu\nu\rho\sigma}\partial_\sigma\phi
\label{solPitophi}
\ee
with $\phi$ a scalar field, according to the Poincar\'{e} lemma (assuming no topological obstructions). If we substitute this solution into the action we obtain
\be
\mS=-\frac12\int\dd^4x\sqrt{-g}\partial_\mu\phi\partial^\mu\phi\,
\ee
that shows the well-known duality between a Kalb-Ramond field and a massless free scalar field that can be found in numerous places in the literature. Less common is to find this duality for the case of a self-interacting Kalb-Ramond field,\footnote{See however e.g. \cite{Horn:2015zna} where the duality for a self-interacting Kalb-Ramond fields is constructed in the context of an EFT description of super-fluids.} although the extension is straightforward. Thus, we will explicitly show in the following that the self-interacting Kalb-Ramond field described by the action
\be
\mS=\int \dd^4 x \sqrt{-g}F(Y)
\label{2form-k}
\,,
\ee
is dual to a $K$-essence,
\be
\mS_{\rm dual}=\int \dd^4 x \sqrt{-g}\mP(X)\,, 
\label{scalar-k}
\ee
provided the following conditions of non-degeneracy of the dual transformation holds:
\be
 F_Y\neq 0 \,, \quad F_Y+2YF_{YY}\neq 0
 \,, \quad
 \mP_X\neq 0 \,, \quad \mP_X+2X\mP_{XX}\neq 0 
\,,
\label{regular_con}
\ee
where $F$ and $\mP$ are respectively functions of $X=-\frac{1}{2}(\partial \phi)^2$ and $Y=- \frac{1}{12} H_{\mu\nu\rho}H^{\mu\nu\rho} $. The on-shell relations between the two dual descriptions are given by
\be
\mP=F-2YF_Y\,, \quad F_Y H_{\mu\nu\rho}= \varepsilon_{\mu\nu\rho\sigma}\partial^\sigma \phi 
\,,
\label{P_rel}
\ee
and
\be
F=\mP-2X\mP_X\,, \quad \mP_X \partial_{\mu}\phi=\frac{1}{6} \varepsilon_{\mu\nu\rho\sigma}H^{\nu\rho\sigma}
\,,
\label{F_rel}
\ee
Furthermore, we can use \eqref{P_rel} and \eqref{F_rel} to obtain the relations
\be
\frac{\dd X}{\dd Y}= -\frac{1}{\mP_X (\mP_X+2X\mP_{XX})} 
\,, \quad
\frac{\dd Y}{\dd X}=-\frac{1}{F_Y(F_Y+2YF_{YY})}
\,,
\ee
and
\be
F_Y=\frac{1}{\mP_X}\,, \quad
F_Y+2Y F_{YY}=\frac{1}{\mP_X+2X\mP_{XX}}
\,,
\label{dual_relation}
\ee
which directly show that the conditions \eqref{regular_con} are the regularity conditions of the dualisation. We also note that the energy-momentum tensors
\bea
T^{\mu\nu}&=&\frac{1}{2}F_Y H^{\mu\alpha\beta}H^{\nu}{}_{\alpha\beta}+ F g^{\mu\nu}
\nonumber \\
&=&F_Y \dualH^{\mu}\dualH^{\nu}+(F-2YF_Y)g^{\mu\nu}
\,, \label{2form-T} 
\\
T_{\rm dual}^{\mu\nu}&=&\mP_X \partial^{\mu}\phi \partial^{\nu}\varphi + \mP g^{\mu\nu}
\,, \label{scalar-T} 
\eea
agree with each other on-shell where $\dualH^{\mu}\equiv \frac{1}{6}\varepsilon^{\mu\nu\rho\sigma}H_{\nu\rho\sigma}$ is the Hodge dual of the field strength.

Let us then proceed to show the explicit construction of the dualisation. We start from the first order formulation of the self-interacting Kalb-Ramond field given by
\be
\mS=\int\dd^4x\sqrt{-g}\left[F(Y_P)-\frac{1}{6}F_Y(Y_P) P^{\mu\nu\rho}(H_{\mu\nu\rho}-P_{\mu\nu\rho})\right]\,, \quad Y_P\equiv -\frac{1}{12}P_{\mu\nu\rho}P^{\mu\nu\rho}
\ee
with $P_{\mu\nu\rho}$ an auxiliary field. We can reach this first order form from the general second order formulation described by \eqref{2form-k} as usual by performing the Legendre transformation. It is convenient to use $\dualP_{\mu}\equiv \frac{1}{6}\varepsilon^{\mu\nu\rho\sigma}P_{\nu\rho\sigma}$ to see the invertibility condition of the Legendre transformation. The Lagrangian is
\be
\mS=\int\dd^4x\sqrt{-g}\Big[F(Y_P)+F_Y(Y_P) \dualP_{\mu}(\dualH^{\mu}-\dualP^{\mu})\Big]\,, \quad Y_P= \frac{1}{2}\dualP_{\mu} \dualP^{\mu}
\,.
\ee
The variation w.r.t.~the auxiliary variable yields
\be
\Big( F_Y g_{\mu\nu}+F_{YY}\tilde{P}_{\mu}\tilde{P}_{\nu} \Big)(\dualH^{\mu}-\dualP^{\mu})=0
\,,
\ee
which is solved by $\dualP^{\mu}=\dualH^{\mu}$ as long as the matrix $F_Y g_{\mu\nu}+F_{YY}\dualP_{\mu}\dualP_{\nu}$ is invertible. Therefore, the invertibility condition of the Legendre transformation is ${\rm det}(F_Y g_{\mu\nu}+F_{YY}\dualP_{\mu}\dualP_{\nu})\neq 0$. This determinant can be easily computed to be 
\be
{\rm det}(F_Y g_{\mu\nu}+F_{YY}\dualP_{\mu}\dualP_{\nu})=F_Y^3(F_Y+2YF_{YY})
\ee
so the invertibility is guaranteed under the following conditions:
\be
F_Y \neq 0 \,, \quad F_Y+2YF_{YY}\neq 0
\,.
\label{invert_F}
\ee 
Upon the field redefinition
\be
F_Y P^{\mu\nu\rho}\equiv  \Pi^{\mu\nu\rho}
\label{eq:38}
\ee
that is invertible provided \eqref{invert_F} holds so we can express $P^{\mu\nu\rho}$ in terms of $\Pi^{\mu\nu\rho}$,  we arrive at the first order formulation
\be
\mS=\int\dd^4x\sqrt{-g}\left[-\frac12 \Pi^{\mu\nu\rho} \partial_{[\mu} B_{\nu\rho]} + \mH(Y_{\Pi})\right],
\quad Y_{\Pi}\equiv -\frac{1}{12}\Pi_{\mu\nu\rho}\Pi^{\mu\nu\rho}
\,,
\ee
where
\be
\mH(Y_{\Pi})=\Big[F(Y_P)-2Y_P F_Y(Y_P) \Big]_{Y_P=Y_P(Y_{\Pi})}\,.
\label{HToF}
\ee
The field equation for $\Pi$ is now modified to
\be
H_{\mu\nu\rho}=- \mH_{Y_{\Pi}} (Y_{\Pi})\, \Pi_{\mu\nu\rho}
\label{eq:Pi1}
\ee
that affects the relation between the field strength $H_{\mu\nu\rho}=3\partial_{[\mu} B_{\nu\rho]}$ and the conjugate momentum $\Pi_{\mu\nu\rho}$, where $\mH_{Y_{\Pi}}=\dd \mH/\dd Y_{\Pi}$.  On the other hand, the equation for the 2-form remains the same so we can still write
\be
\Pi_{\mu\nu\rho}=\varepsilon_{\mu\nu\rho\sigma}\partial^\sigma\phi\,.
\ee
From the square of this expression we obtain $Y_{\Pi}=-X$ and then the dual action in terms of $\phi$ is
\be
\mS_{\rm dual}=\int\dd^4x\sqrt{-g}\mH(-X)\equiv\int\dd^4x\sqrt{-g}\mP(X)
\,.
\ee
The on-shell relation between the scalar field and the 2-form field is
\be
H_{\mu\nu\rho}=\mP_X(X) \varepsilon_{\mu\nu\rho\sigma}\partial^\sigma \phi
\,.
\label{duality1}
\ee
whose square leads to
\be
Y=-\mP_X^2(X) X
\,.
\ee
We notice that \eqref{eq:38} and $P_{\mu\nu\rho}=H_{\mu\nu\rho}$ provide another on-shell relation
\be
F_YH_{\mu\nu\rho}= \varepsilon_{\mu\nu\rho\sigma}\partial^\sigma \phi 
\quad \implies \quad
YF_Y^2(Y)=-X
\,,
\label{duality2}
\ee
that allows to express $Y$ in terms of $X$ for a given $F(Y)$.
The relation between the particular $\mP(X)$ theory and $F(Y)$ can be obtained from the relation between $\mH$ and $F$ deduced in \eqref{HToF} so we have
\be
\mP(X)=F(Y(X))-2Y(X) F_Y(Y(X))
\,.
\label{dualPF}
\ee

We then study the inverse transformation. As before, we can proceed by performing a Legendre transformation that recasts the action into the first order formalism as follows
\be
\mS_{\rm dual}=\int \dd^4x \sqrt{-g}\Big[ \mP (X_p) - \mP_X(X_p) p^{\mu} (\partial_{\mu}\phi-p_{\mu}) \Big]
, \quad X_p\equiv -\frac{1}{2}p_{\mu}p^{\mu}
\,,
\label{scalar-k2}
\ee
where $p_{\mu}$ is an auxiliary vector field. The variation w.r.t.~$p_{\mu}$ yields
\be
(\mP_X g_{\mu\nu}-\mP_{XX}p_{\mu}p_{\nu}) (p^{\mu}-\partial^{\mu}\phi)=0
\,,
\label{eom_p}
\ee
so \eqref{scalar-k2} is equivalent to the $K$-essence described by \eqref{scalar-k} as long as ${\rm det}(\mP_X g^{\mu\nu}-\mP_{XX}p^{\mu}p^{\nu})\neq 0$, i.e.,~
\be
\mP_X\neq 0 \,, \quad \mP_{X}+2X\mP_{XX} \neq 0
\,.
\label{invert_P}
\ee
Let us introduce the new variable $\pi^{\mu}$ under the relation
\be
\pi^{\mu}= \mP_X p^{\mu}
\,,
\label{Pip_rel}
\ee
which can be solve by $p_{\mu}=p_{\mu}(\pi_{\alpha})$ under \eqref{invert_P} according to the implicit function theorem.
We thus obtain the first order form of the $K$-essence,
\begin{align}
\mS_{\rm dual}=\int \dd^4 x \sqrt{-g}\left[ - \pi^{\mu}\partial_{\mu} \varphi + \mathscr{H}(X_{\pi})\right]
\,, \quad
X_{\pi}= -\frac{1}{2}\pi^{\mu}\pi_{\mu}
\,,
\label{scalar-k3}
\end{align}
where
\begin{align}
\mathscr{H}(X_{\pi})
=\Big[ \mP(X_p)-2X_p \mP_X (X_p) \Big]_{X_p=X_p(X_{\pi}) } 
\,.
\end{align}
The equation for $\phi$ now says that $\partial_\mu \pi^\mu=0$ so we can write the momentum as $\pi^\mu=\frac16 \epsilon^{\mu\nu\rho\sigma} \partial_{[\nu} B_{\rho\sigma]}$ that we can use to integrate $\pi^{\mu}$ out and write the dual action
\begin{align}
\mS=\int \dd^4 x \sqrt{-g}\mathscr{H}(-Y) \equiv \int \dd^4 x \sqrt{-g} F(Y)
\,,
\end{align}
with the on-shell relations \eqref{F_rel}. Hence, we have established the duality between the self-interacting Kalb-Ramond field and the $K$-essence under the conditions \eqref{regular_con}.

This duality is precisely what permits to establish a duality relation between the cosmology of a shift-symmetric scalar field with a time-dependent profile and the cosmological solutions presented above for a Kalb-Ramond field. Let us see how this works more explicitly. For a time-dependent scalar field profile $\phib(t)$ in a FLRW universe described by the line element 
\be
\dd s^2=-\bar{N}^2(t)\dd t^2 + a^2(t) \dd \vec{x}^2\,,
\ee 
the duality \eqref{duality1} gives
\be
H_{ijk}=\mP_X(\bar{X})\sqrt{-g}\epsilon_{ijk}\partial^0\phib.
\ee
where we have used the relation $\varepsilon_{ijk0}=\sqrt{-g}\epsilon_{ijk}$ for $\varepsilon_{0123}=\sqrt{-g}$ and $\epsilon_{123}=+1$. This means that the field strength of the two form must take the form $H_{ijk}=B\epsilon_{ijk}$ with $B$ satisfying
\be
B=a^3 \mP_X \frac{\partial_0\phib}{\bar{N}}\,.
\ee
The equation of motion of the $K$-essence yields
\be
\partial_0\left(a^3 \mP_X \frac{\partial_0\phib}{\bar{N}}\right) = 0
\,,
\ee
meaning that $B$ is nothing but a constant of motion.
This is precisely the 2-form configuration \eqref{Hconfig}. Notice that this relation is invariant under time-reparameterisations, as it should because the 2-form configuration does not break time diffeomorphisms. Thus, we have seen that the non-trivial realisation of homogeneity and isotropy with the inhomogeneous 2-form configuration is dual to the usual cosmological solutions for a shift-symmetric scalar field theory.

Let us rewrite the on-shell relations \eqref{P_rel} and \eqref{F_rel} into
\bea
\star(F_Y \dd {\bf B}) &=&  \dd \phi \,, \\
\star(\mP_X \dd \phi) &=& \dd {\bf B}
\,, 
\label{PXdp=dB}
\eea
where $\star$ is the Hodge star operator and $ {\bf B}=\frac{1}{2}B_{\mu\nu}\dd x^{\mu} \wedge \dd x^{\nu}$. Since we have the identity $\dd^2 =0$, we obtain
\bea
\delta(F_Y \dd {\bf B}) &=& 0
\,, \\
\delta(\mP_X \dd \phi) &=& 0
\,,
\eea
which agree with the equations of motion of the self-interacting Kalb-Ramond field and the $K$-essence, respectively. A special case of this relation has been discussed above to show that $B$ is a constant of motion. This provides not only a consistency check of the dualisation but also shows that the identity in one side is mapped into the field equation in the dual side, which is the usual property under dualisation in the absence of matter sources.

Although the correspondence of the energy-momentum tensors has been already confirmed explicitly, it is useful to see how the corresponding energy-momentum tensors are related in a more formal way.  For that, let us notice that the duality can be regarded as just a (non-local) field redefinition $B_{\mu\nu}(x)=B_{\mu\nu}[\phi,g_{\alpha\beta}](x)$. The energy-momentum tensors are then related by
\be
\frac12\sqrt{-g} T^{\mu\nu}_{\rm dual}(x)=\left(\frac{\delta \mS[g,B]}{\delta g_{\mu\nu}(x)}\right)_{B}+\int d^4y \frac{\delta \mS[g,B]}{\delta B_{\alpha\beta}(y)}\frac{\delta B_{\alpha\beta}[\phi,g_{\alpha\beta}](y)}{\delta g_{\mu\nu}(x)}
\ee
where the first term is the energy-momentum tensor for the 2-form (it is computed by keeping $B_{\alpha\beta}$ constant) and the second term is generated from the metric-dependence of the duality relation. However, since this term is proportional to the field equation of motion, we have that the two energy-momentum tensors coincide on-shell as long as $\delta B_{\alpha\beta}[\phi,g_{\alpha\beta}](y)/\delta g_{\mu\nu}(x)$ is well defined, i.e.,~the field redefinition is regular.

As an interesting remark, we could notice that the shift symmetry of the scalar field in the dual description of the theory must be exact and cannot be softly broken (unless we introduce non-local operators in the 2-form side which we will discuss later). This is so because it derives from the dual 2-form theory with an exact gauge symmetry that ultimately forces the shift symmetry. Were we to introduce an explicit gauge breaking term in the action for the 2-form (e.g. a mass term), this would result in an enhanced phase space with two additional propagating dof's. The dual of this theory is obviously not a non-symmetric scalar field (that would preserve the number of dof's) but it is a Proca field, that correctly matches the number of dofs (see Sec.~\ref{sec:massive2form} below for the explicit dualisation). From this more general case, we can nicely recover the shift symmetry. If we take an appropriate decoupling limit where we send the mass of the Kalb-Ramond field to zero (and any other coupling constant associated to higher order operators), the dual description is then a vector gauge theory plus the decoupled longitudinal mode that has the usual shift symmetry for Goldstone bosons. Since the original symmetry is an exactly realised gauge symmetry, we cannot get any pseudo-Goldstone boson where the soft breaking of the shift symmetry (that is related to the very existence of a quasi de Sitter solution with an approximate scale invariance) is the responsible for the eventual exit from the inflationary phase. This is not a viable mechanism in the present scenario because of the gauge origin of the shift symmetry so the end of inflation and the reheating period must occur from a different mechanism.\footnote{Let us note in passing that the Swampland conjecture about the impossibility of having exact global symmetries in the low energy effective field theory of a string theory unless they originate from gauge symmetries is nicely realised in the Kalb-Ramond cosmology: The exact shift symmetry of the scalar field is simply a consequence of the gauge symmetry in the dual description.} A possibility would be to have a soft breaking of the gauge symmetry so that inflation would end when the additional polarisations of the massive Kalb-Ramond field become relevant. Let us note however that the pure massless Kalb-Ramond field with self-interactions cannot support slow roll inflationary solutions because those solutions are destabilised by a Laplacian instability, as we will discuss in Sec.~\ref{Sec:CosmologyKalbRammond}.

An interesting property of the duality relation between the 2-form and the scalar field is the different (dual) realisations of the cosmological principle. On the 2-form side, time translations are not broken and therefore one natural question to posse is what determines the end of inflation or, in other words, at what time we are to match the correlation functions that will give the initial conditions for the post-inflationary era. This issue also arises in models like solid inflation \cite{Endlich:2012pz} where the symmetries of the background configuration also preserve time reparameterisations invariance. If we now go to the shift symmetric scalar side, the scalar field takes a time-dependent profile, thus explicitly breaking time diffs and giving a natural cosmic clock variable. Let us notice however that, due to the shift symmetry, there is a combination of time translations and shifts that serve as a diagonal time translational invariance.

Before closing this subsection, we shall argue a possible way of the shift symmetry breaking by the use of non-local operators in the 2-form side. We should recall that non-local operators can secretly describe additional degrees of freedom. Let us suppose that the shift symmetry breaking is small enough to use the relation \eqref{duality2} at the leading-order. Then, we can formally obtain
\be
\phi = \int \dd \phi=\int \star(F_Y \dd {\bf B})
\,.
\ee 
The soft breaking of the shift symmetry necessarily leads to a non-local operator in the dualised 2-form theory. Appearance of non-locality in the action would imply that a degree of freedom that couples to the 2-form has been integrated out. Note that in the comoving gauge of the scalar field, also called the unitary gauge, the scalar field is identified with the time coordinate $t$. Hence, the soft breaking of the shift symmetry in the comoving gauge may be achieved by adding $t$ to the dualised theory at least at the leading-order.

\subsection{Duality as a weak/strong duality}
\label{sec:weak/strong}
An important property of the duality is that an originally strongly coupled theory may appear as a weakly coupled theory in the dual formulation. In this subsection, we will show how this is indeed the case in our duality between the $K$-essence and the self-interacting Kalb-Ramond field. 

Let us first investigate the cubic interactions of the $K$-essence and the self-interacting Kalb-Ramond field. For simplicity, we ignore gravity and consider perturbations around the cosmological configurations, $\phi \propto t$ for the $K$-essence and $B_{ij} \propto \epsilon_{ijk}x^k$ for the Kalb-Ramond field. The Lagrangian of the $K$-essence up to the cubic order in perturbations is
\begin{align}
\delta \mL_{\rm dual} &=\frac{\mP_X}{2c_s^2}[\dot{\pi} -c_s^2 (\partial_i \pi)^2] + \frac{2\lambda_X}{(2X)^{3/2}}\dot{\pi}^3 - \frac{\mP_X(1-c_s^2)}{2 (2X)^{1/2} c_s^2 }\dot{\pi}(\partial_i \pi)^2
\,,
\end{align}
where $\pi$ is the perturbation of the scalar field and $\lambda_X \equiv X^2\mP_{XX}+\frac{2}{3}X^3\mP_{XXX}$. We rescale the time $t \to t/c_s$ to find a relativistic kinetic term and canonically normalise the field $\pi \to (c_s/\mP_X)^{1/2}\pi$. The action is then
\begin{align}
\delta \mS_{\rm dual}=\int \dd^4x \delta \mL_{\rm dual} \to \int \dd^4x \left\{ \frac{1}{2}[\dot{\pi} - (\partial_i \pi)^2] + \frac{2c_s^{7/2} \lambda_X}{(2X\mP_X)^{3/2}} \dot{\pi}^3 - \frac{1-c_s^2}{2c_s^{1/2} (2X\mP_X)^{1/2}} \dot{\pi} (\partial_i \pi)^2 \right\}
\,.
\end{align}
As for the 2-form field, we shall focus on the scalar-type of perturbations for which the 2-form field takes the form
\begin{align}
B_{0i}=0\,, \quad B_{ij}=\epsilon_{ijk}\left(\frac{1}{3}Bx^k + \partial^k \varpi\right)
\end{align}
with the perturbation $\varpi$. Here, we have used the gauge freedom to set $B_{0i}=0$. Furthermore, we are neglecting the non-dynamical perturbation $\varpi^i=\epsilon^{ijk}B_{jk}$ that is not expected to play a relevant role for the subsequent analysis regarding the strong coupling scale obtained from the cubic interactions. The perturbed Lagrangian of the self-interacting Kalb-Ramond field is given by
\begin{align}
\delta \mL = \frac{F_Y}{2}\Big[ (\partial_i \dot{\varpi})^2- c_s^2 (\Delta \varpi)^2\Big] + \frac{2\lambda_Y}{(-2Y)^{3/2}}(\Delta \varpi)^3
-\frac{F_Y(1-c_s^2)}{2(-2Y)^{1/2}}(\Delta \varpi)(\partial_i \varpi)^2
\end{align}
where $\Delta \equiv \partial^i \partial_i$ and $\lambda_Y \equiv Y^2F_{YY}+\frac{2}{3}Y^3F_{YYY}$. As in the case of the $K$-essence, we normalise the time and the field according to $t \to t/c_s$ and $\varpi \to \varpi/(c_s F_Y)^{1/2}$, yielding
\begin{align}
&\delta \mS = \int \dd^4x \delta \mL 
\\ \nonumber
&\to \int \dd^4 x \left\{ \frac{1}{2}[(\partial_i \dot{\varpi})^2- (\Delta \varpi)^2] 
+\frac{2\lambda_Y}{c_s^{5/2}(-2YF_Y)^{3/2}}(\Delta \varpi)^3 
-\frac{1-c_s^2}{2c_s^{1/2}(-2YF_Y)^{1/2}}(\Delta \varpi)(\partial_i \dot{\varpi})^2
\right\}
\,.
\end{align}
Note that the canonical normalisation further requires the replacement $\varpi \to \varpi/k$ with $k$ being the momentum but we shall not perform it explicitly since the structure of the cubic interactions does not change.
By using the duality relations $X\mP_X = -YF_Y$ and $\lambda_Y=c_s^6 \lambda_X$ which are derived from \eqref{P_rel}-\eqref{dual_relation}, we finally obtain
\begin{align}
&\delta \mS = \int \dd^4 x \left\{ \frac{1}{2}\left[(\partial_i \dot{\varpi})^2- (\Delta \varpi)^2\right] 
+\frac{2c_s^{7/2}\lambda_X}{(2X\mP_X)^{3/2}}(\Delta \varpi)^3 
-\frac{1-c_s^2}{2c_s^{1/2}(2X\mP_X)^{1/2}}(\Delta \varpi)(\partial_i \dot{\varpi})^2
\right\}
\,.
\end{align}
Therefore, the cubic interactions of both the scalar and the 2-form descriptions have the same scaling behaviour and we find that the theory is strongly coupled as $c_s \to 0$ in both formulations. However, as is well known, higher-derivative corrections become important in the limit $c_s \to 0$ and we have to take them into account to properly study the limit $c_s\to 0$~\cite{Arkani-Hamed:2003pdi,Arkani-Hamed:2003juy}. 

It is important to notice at this point that the duality for higher order derivative terms cannot be established as straightforwardly as for the leading order terms involving only first derivatives. The reason can be understood easily by noticing that a correction $\Delta\Lag=\frac{1}{\Lambda^2}(\Box\phi)^2$ to the $K$-essence Lagrangian effectively introduces a scalar ghost of mass $\Lambda^2$ and this ghost field cannot be dualised to a 2-form. On the other hand, a correction $\Delta\Lag=\frac{1}{\Lambda^2}(\partial_\mu H^{\mu\nu\rho})^2$ gives rise to a 2-form ghost of mass $\Lambda^2$, whose dual is a 1-form, not a scalar (see below Sec. \ref{sec:massive2form}). Thus, we conclude that the higher order corrections will typically introduce different numbers of dof's so the duality cannot be established. This is shown in more detail in Appendix \ref{sec:AppHO}. Obviously, from the EFT perspective there will be an infinite number of terms with arbitrarily higher derivatives, but they all must be treated perturbatively so the new would be degrees of freedom will never appear below the cut-off scale of the EFT. However, we will see that this difference in dof's from the higher order corrections can have an impact on the cut-off scale for the two dual formulations. In particular, we will see that one formulation can admit a consistent truncation while in the dual formulation the full hierarchy of higher order derivatives might be necessary. This is an important feature that clearly shows how working in one formulation can be more advantageous than its dual formulation from the EFT perspective.

To understand the duality even in the presence of such higher-derivative corrections, we shall consider a (partial) UV completion of the $K$-essence/self-interacting Kalb-Ramond field by following \cite{Babichev:2016hys,Babichev:2017lrx,Babichev:2018twg,Mizuno:2019pcm,Mukohyama:2020lsu,Aoki:2021ffc}. Here, we adopt the simplest partial UV completion with the help of a heavy field $\chi$. The Lagrangian is given by
\be
\mL_{\rm UV}=-\frac{1}{2}\gamma_{AB}(\chi) \partial_{\mu}\Phi^A\partial^{\mu}\Phi^B - V(\chi)\,, \quad \Phi^A=(\chi,\phi)
\,,
\label{NLSM}
\ee
where the field-space metric $\gamma_{AB}(\chi)$ and the potential $V(\chi)$ are supposed to be independent of $\phi$ to respect the shift symmetry of $\phi$. As we will explain shortly (see \cite{Aoki:2021ffc} for more details), the $K$-essence arises as the leading-order EFT of the non-linear sigma model described by \eqref{NLSM}. Note that \eqref{NLSM} may be also an EFT of a more fundamental theory. However, the applicable range of \eqref{NLSM} is wider than that of the $K$-essence and this is sufficient for the present purpose.

In the two-field model, the freedom of the field redefinition allows us to set the field-space metric to be $\gamma_{AB}={\rm diag}[1,~f(\chi)]$ with a positive function $f(\chi)$. The Lagrangian is then
\be
\mL_{\rm UV}=-\frac{1}{2}(\partial \chi)^2 + f(\chi) X - V(\chi)
\,,
\label{UV_scalar}
\ee
and the field equation for $\chi$ is given by
\be
\Box \chi + X f_{\chi}-V_{\chi}=0
\,. 
\label{eom_chi}
\ee
We solve \eqref{eom_chi} by treating the derivatives as perturbations while keeping the non-linearity of $X$; that is, we assume that $\partial^n \chi$ and $\partial^{n+1}\phi$ with $n\geq 1$ are small quantities compared with a mass scale $M^2$ that we will introduce shortly whereas we do not assume the smallness of $\partial \phi$. The solution of $\chi$ is found to be
\be
\chi=\chi_0(X) + \mathcal{O}(\partial^2/M^2)
\ee
where $\chi_0(X)$ is a root of
\be
Xf_{\chi}-V_{\chi}=0
\,,
\label{eomleading_chi}
\ee
and
\be
M^2 \equiv \Big[ V_{\chi\chi}-X f_{\chi\chi} \Big]_{\chi=\chi_0(X)}
\,.
\ee
The condition $M^2 \neq 0$ guarantees the existence of the root $\chi_0(X)$ at least locally. After integrating out $\chi$, the effective Lagrangian of $\phi$ is given by
\be
\mL_{\rm EFT}=\mP(X)-\frac{\mP_{XX}}{2M^2}(\partial X)^2 + \mathcal{O}(\partial^4/M^4)
\,,
\label{LEFT_scalar}
\ee
with
\be
\mP(X)\equiv f(\chi_0(X))X-V(\chi_0(X))
\,,
\ee
up to the sub-leading order of the derivative expansion. Here, to get the expression of the second term of \eqref{LEFT_scalar}, we have used
\be
\frac{\dd \chi_0}{\dd X} = \frac{f_{\chi}}{M^2}
\,,
\ee
and then
\be
\mP_X=f\,, \quad \mP_{XX}=\frac{f_{\chi}^2}{M^2}
\,,
\label{PX_rel}
\ee
where the functions on r.h.s.~are evaluated at $\chi=\chi_0(X)$. 

Next, we consider the dual of the UV Lagrangian \eqref{UV_scalar}. The first order form of \eqref{UV_scalar} is
\be
\mL_{\rm UV}=-\frac{1}{2}(\partial \chi)^2 +\frac{1}{2} f^{-1}(\chi) \pi_{\mu}\pi^{\mu} - V(\chi) - \pi^{\mu}\partial_{\mu} \phi
\,,
\ee
with $\pi_{\mu}$ an auxiliary variable. The field equation for $\phi$ is solved by $\pi^{\mu}=\frac{1}{6}\varepsilon^{\mu\nu\rho\sigma}H_{\nu\rho\sigma}$, yielding the dualised Lagrangian
\be
\mL_{\rm UV}=-\frac{1}{2}(\partial \chi)^2 + f^{-1}(\chi) Y- V(\chi) 
\,,
\label{UV_2form}
\ee
in terms of the 2-form field. The on-shell relation between the scalar field and the 2-form field is
\be
Y=-f^2(\chi)X
\,.
\label{onshell_UV}
\ee
We then solve the field equation for $\chi$ by taking the derivative expansion. The solution is 
\be
\chi = \chi_0'(Y) + \mathcal{O}(\partial^2/\mM^2)
\,,
\ee
with the root of
\be
f_{\chi}(\chi'_0)\frac{Y}{f^2(\chi'_0)}+V_{\chi}(\chi'_0)=0
\,,
\ee
where the mass scale associated with the derivative expansion is now given by
\be
\mM^2 \equiv \left[ V_{\chi\chi} + \frac{Y}{f^2} f_{\chi\chi}- 2 \frac{Y}{f^2} \frac{f_{\chi}^2}{f} \right]_{\chi=\chi_0'(Y)}
\,.
\ee
Again, the existence of the root $\chi_0'(Y)$ is guaranteed by $\mM^2\neq 0$. The resultant effective Lagrangian in terms of the 2-form field is 
\be
\mL_{\rm EFT}=F(Y)-\frac{F_{YY}}{2\mM^2}(\partial Y)^2 + \mathcal{O}(\partial^4/\mM^4)
\,,
\label{LEFT_2form}
\ee
with
\be
F(Y) \equiv f^{-1}(\chi_0'(Y)) Y- V(\chi_0'(Y))
\,.
\ee 
Similarly to \eqref{LEFT_scalar}, we have used
\bea
\frac{\dd \chi'_0}{\dd Y}&=& -\frac{f_{\chi}}{f^2 \mM^2}
\,, \\
F_Y&=&f^{-1}\,, \quad F_{YY}=\frac{f_{\chi}^2}{f^4\mM^2}
\,.
\eea
Note that the relation between $P(X)$ and $F(Y)$ coincides with that we have established in the previous subsection at the leading-order of the derivative expansion. 

We emphasise that the expansion parameters of the derivative expansion are $\partial/M$ in \eqref{LEFT_scalar} and $\partial/\mM$ in \eqref{LEFT_2form}, respectively.
The scale $\mM$ coincides with $M$ around the trivial background $X=0=Y$, while they denote different scales when the fields have non-trivial background values. For simplicity, we consider the background with $X,Y=$ constant and evaluate $M^2$ and $\mM^2$ on this background. The on-shell relation \eqref{onshell_UV} concludes $\chi_0(X)= \chi_0'(Y)=$ constant and
\be
\mM^2 = V_{\chi\chi} - f_{\chi\chi}X+2X \frac{f_{\chi}^2}{f} = M^2 +2X \frac{f_{\chi}^2}{f} 
\,,
\label{Mcal_M}
\ee
evaluated at $X,Y=$ constant.
Since the ghost-free condition of the UV theory requires $f>0$, we find
\be
\begin{cases}
\mM^2 > M^2\,, & {\rm for}~~X>0,~Y<0 \,, \\
\mM^2 = M^2\,, & {\rm for}~~X=0,~Y=0 \,, \\
\mM^2 < M^2\,, & {\rm for}~~X<0,~Y>0 \,. \\
\end{cases}
\ee
Therefore, the convergence of the derivative expansion depends on the description of the theory and the property of the background. To make the discussion concrete, we focus on the cosmologically relevant case $X>0$ for a while. By using \eqref{PX_rel} and \eqref{Mcal_M}, we obtain
\be
c_s^2 \mM^2=M^2
\,,
\ee
where $c_s^2=\mP_X/(\mP_X+2X\mP_{XX})=(F_Y+2Y F_{YY})/F_Y$ is the sound speed of the perturbations around the background with $X>0,~Y<0$. The hierarchy $\mM^2 \gg M^2$ is realised in a largely Lorentz-violating background, $c_s^2\ll 1$, and this is indeed what we are interested in. The derivative expansion in the $K$-essence description may converge at $\partial/M \ll 1$ while the convergence in the 2-form description only requires $\partial/\mM=c_s \partial/M \ll 1$. The difference between $M^2$ and $\mM^2$ are examined in~\cite{Aoki:2021ffc} by studying the dispersion relation of the two-field model. Although the scale $M^2$ appears as the scale associated with the derivative expansion in \eqref{LEFT_scalar}, $M^2$ does not agree with the actual mass scale of the heavy mode around the $X>0$ background. The actual mass scale is given by $\mM^2$, so one can extend the validity of the single-field EFT even beyond $M^2$ if one considers a resummation of the higher-derivative terms. Here, we have found that the dualisation of the $K$-essence indeed provides such a resummation. In other words, at least in this UV completion, the resultant EFT in terms of the 2-form admits a weakly coupled description with a finite number of higher-derivative operators while the $K$-essence description requires an infinite number of higher-derivative operators in the limit $c_s \to 0$.

The situation is opposite in $X<0,~Y>0$. The scale $M^2$ is larger than $\mM^2$ and the actual mass scale of the heavy mode is given by $M^2$~\cite{Aoki:2021ffc}. Although the 2-form description becomes strongly coupled at $\partial/\mM =\mathcal{O}(1)$, the $K$-essence provides a resummation of the strongly coupled 2-form field and the single-field description is valid as long as $\partial/M \ll 1$.

It is worth mentioning that a violation of the regularity condition \eqref{regular_con} may be interpreted as an infinitely strong coupling in one side of the duality while the other side might be oblivious to that. The simplest example is found from a massive $U(1)$ scalar field with a quartic self-interaction:
\be
\mL=-\frac{1}{2}|\partial \Phi|^2 -\frac{M_{\Phi}^2}{2}|\Phi^2| -\frac{\lambda}{4}|\Phi|^4
\,.
\label{U(1)}
\ee
In terms of $\chi$ and $\phi$ defined by $\Phi=\chi e^{i\phi}$, the Lagrangian is given by
\be
\mL=-\frac{1}{2}(\partial \chi)^2+\chi^2 X - \frac{M_{\Phi}^2}{2}\chi^2 - \frac{\lambda}{4}\chi^4
\,,
\ee
which corresponds to $f=\chi^2$ and $V=\frac{M_{\Phi}^2}{2}\chi^2+\frac{\lambda}{4}\chi^4$. One then obtains
\be
M^2=2\lambda \chi_0^2\,, \quad \chi_0=\sqrt{\frac{2X-M_{\Phi}^2}{\lambda}}
\,,
\ee
and
\be
\mP(X)=\frac{1}{\lambda}(X-M_{\Phi}^2/2)^2
\,.
\ee
We then consider the limit $\lambda \to 0$ while keeping $\chi_0$ finite which is nothing but the weak coupling limit from the point of view of the $U(1)$ scalar. However, the $K$-essence description is infinitely strongly coupled (that is, we cannot truncate the derivative expansion at a finite order) because of $M^2\to 0$. On the other hand, when we consider the dual description,
\be
\mL=-\frac{1}{2}(\partial \chi)^2+\chi^{-2} Y - \frac{M_{\Phi}^2}{2}\chi^2 - \frac{\lambda}{4}\chi^4
\,,
\ee
we find 
\bea
 \chi_0'{}^2&=&\frac{1}{M_{\Phi}}\sqrt{-2Y}+\frac{\lambda}{M_{\Phi}^4}Y+\mathcal{O}(\lambda^2)
 \,, \\
F(Y)&=&-M_{\Phi} \sqrt{-2 Y} + \frac{\lambda}{2M^2_{\Phi}} Y+\mathcal{O}(\lambda^2)
\,,
\label{F_freeU(1)}
\eea
and
\be
\mM^2=(2M_{\Phi})^2+\mathcal{O}(\lambda) \,.
\ee
Hence, the derivative expansion in the dual description can be truncated at a finite order even in $\lambda \to 0$ thanks to a finite $\mM^2$. The functional form \eqref{F_freeU(1)} leads to
\be
F_Y+2YF_{YY}=\mathcal{O}(\lambda)
\,,
\ee
so the limit $\lambda \to 0$ indeed corresponds to the singular point of the duality. 

Needless to say, the discussion so far is based on a simple (partial) UV completion of the $K$-essence/self-interacting Kalb-Ramond field, but the two-field model is not the unique UV completion. From the low-energy perspective, what we know is that the duality is singular when at least one of \eqref{regular_con} is violated. As we have argued in this subsection, such a singular point is a singularity of one description while the theory in the dual side may be weakly coupled thanks to appropriate higher-derivative corrections. We will discuss these singular examples in the next section. Here, we emphasise that our strong/weak duality is not a statement about the $K$-essence and the self-interacting Kalb-Ramond field themselves since both descriptions {\it without derivative corrections} are strongly coupled in the limit $c_s \to 0$. A higher derivative operator is required to cure the strong coupling issue when $c_s \to 0$. However, the duality transformation also affects the structure of higher-derivative corrections and then a theory with a finite number of higher-derivative operators may appear as a theory with an infinite number of higher-derivative operators on the dual side. There are two possibilities to have $c_s\to 0$ around the cosmological background: the first is the limit
\begin{align}
F_Y+2Y F_{YY}=\frac{1}{\mP_X + X \mP_{XX}} \to 0
\label{zerospeed1}
\end{align}
and the second possibility is
\begin{align}
\mP_X =\frac{1}{F_Y} \to 0
\,.
\label{zerospeed2}
\end{align}
In the former case, as we have seen, the 2-form description is well-behaved and an appropriate higher-derivative operator to give a non-relativistic dispersion relation is $(\partial Y)^2$. On the other hand, in the case of \eqref{zerospeed2}, the appropriate description would be the scalar field and the strong coupling is cured by the operator $(\Box \phi)^2$, known as the ghost condensate~\cite{Arkani-Hamed:2003pdi,Arkani-Hamed:2003juy}.

\subsection{Duality as fluids}

The shift symmetric $\mP(X)$ theories can be interpreted as the effective field description of a super-fluid with the shift symmetry related to the internal conserved charge of the fluid. The duality relation between the shift symmetric theories and the 2-form theories is close to the relation between the super-fluid and ordinary fluids (see e.g. ~\cite{Dubovsky:2005xd}). Carter already used a Kalb-Ramond field for the description of perfect fluids in \cite{Carter:1994rv} (see also~\cite{Carter:1995mj}). The dual description of super-fluids in terms of a 2-form field has also been exploited recently to study the interactions of vortex lines and vortex rings from an EFT perspective \cite{Horn:2015zna,Garcia-Saenz:2017wzf}. Although our aim is much more modest than the treatment in those works, it will be instructive to say a few words here on our duality from the fluid perspective to make contact with the results for the multi-Kalb-Ramond case of Sec.~\ref{sec:multiKR}.

A general perfect fluid can be described in terms of three scalar fields $\varphi^a$ ($a=1,2,3$) that represent the comoving coordinates of the fluid. In order to comply with the fluid symmetries, we need to impose an internal shift symmetry and volume-preserving diffeomorphisms: 
\be
\varphi^a\rightarrow \varphi^a +c^a,\quad\text{and}\quad \varphi^a\rightarrow \varphi'{}^a(\varphi^b)\quad\text{with}\quad \det\frac{\partial\varphi'{}^a}{\partial\varphi^b}=1\,.
\ee
The first symmetry reflects the freedom in choosing the spatial origin, as appropriate to describe homogeneous fluids of relevance for cosmologies complying with the cosmological principle. The second symmetry corresponds to our freedom in permuting any two elements of the fluid as long as the volume (or the number density) is kept fixed or, from a more physical point of view, it does not cost energy to move around a fluid element as long as we do not vary its volume. The shift symmetry requires the Lagrangian to contain derivatives of the fields so, at lowest order in derivatives, the fundamental object is $B^{ab}\equiv\partial_\mu\varphi^a\partial^\mu\varphi^b$. Furthermore, the volume-preserving diffeomorphisms are realised if only the determinant of this matrix $B\equiv\det B^{ab}$ is utilised to describe the fluid dynamics. Thus, the Lagrangian will be $\Lag=F(B)$ at the leading-order in the derivative expansion. The four-velocity and the number density of the fluid are respectively given by
\bea
u_f^{\mu} &=& - \frac{1}{6 \sqrt{B} } \epsilon_{ijk} \varepsilon^{\mu\nu\rho\sigma} \partial_{\nu} \varphi^i \partial_{\rho} \varphi^j \partial_{\sigma} \varphi^k
\,, \\
n_f &=& \sqrt{B}
\,,
\eea
which satisfy the off-shell identity
\be
\nabla_{\mu} ( n_f u_f^{\mu} ) = 0
\,,
\ee
that permit to interpret $J_f^{\mu}=n_f u_f^{\mu}$ as the current of fluid particles.

We consider the infinitesimal volume-preserving diffeomorphisms around the ground state of the fluid, $\langle \varphi^i \rangle \propto x^i$. Denoting the perturbations of $\varphi^i$ by $\delta \varphi^i$, the infinitesimal transformation is given by
\be
\delta \varphi^i(t,\vec{x}) \to \delta \varphi^i(t,\vec{x}) + \xi^i(\vec{x})
\label{infvpreserving}
\ee
where the functions describing the volume-preserving diffeomorphisms satisfy the transverse condition
\be
\partial_i \xi^i(\vec{x}) =0 \,.
\ee
Let us then revisit the 2-form theory around the cosmological background. The background configuration of the 2-form is $\langle B_{0i} \rangle =0,~ \langle C^i \rangle \propto x^i$. We can choose the gauge $\delta B_{0i}=0$ by the use of the symmetry of the 2-form in which the dynamical variables of the 2-form are $\delta C^i$. Note that this gauge is not a complete gauge fixing as the residual gauge symmetry is described by
\be
\delta C^i(t,\vec{x}) \to \delta C^i(t,\vec{x}) + \theta^{Ti}(\vec{x}) \quad \text{s.t.} \quad \partial_i \theta^{Ti}(\vec{x}) = 0
\,,
\label{infvpreserving2}
\ee
which is the same form as the infinitesimal volume-preserving diffeomorphisms \eqref{infvpreserving}. Indeed, the gauge transformation of the 2-form can be regarded as a transformation that preserves the number density of the fluids. We introduce the four-velocity
\be
u^{\mu} = -\frac{\dualH^{\mu} }{\sqrt{-2Y} }
\,,
\ee
which is properly normalised, $u_{\mu}u^{\mu}=-1$, provided $Y<0$. From the geometrical identity $\nabla_{\mu} \dualH^{\mu}=0$, we can interpret the quantity
\be
n = \sqrt{-2Y}
\ee
as the number density. Hence, the gauge transformation of the 2-form does not change the number density $n$ and then the volume of the fluid space.

Note that \eqref{infvpreserving2} is just a part of the gauge transformation. The 2-form symmetry is a local symmetry, meaning that the transformation parameter is a function of not only space but also time, in general. The 2-form symmetry is larger than the volume-preserving diffeomorphisms which discriminates the 2-form from the ordinary fluid. In fact, the 2-form is dual to the super-fluid (the shift symmetric scalar field) that is an irrotational fluid. On the other hand, the ordinary fluid generically has a vortex. Hence, it may be interpreted as that the ``localised volume-preserving diffeomorphisms'' (the 2-form gauge symmetry) prohibits the vortex while the global volume-preserving diffeomorphisms accommodate vorticity.

\begin{table}[t]
  \caption{Duality relations from the fluid perspective. Here, we omit the subscript ``dual'' in the last row for simplicity of notation.}
  \label{table:fluidduality}
  \centering
  \begin{tabular}{cccc}
      \hline
       Field & Symmetries & On-shell equations & Off-shell equations \\
       \hline \hline
       \\
       2-form & $B_{\mu\nu} \to B_{\mu\nu}+2\partial_{[\mu}\theta_{\nu]}$ & \begin{tabular}{c} $\omega_{\mu\nu}=0$ \\ $(\rho+p)a_{\mu}+\mathcal{D}_{\mu}p = 0$ \end{tabular} & $\frac{\rm D}{{\rm D}\tau}\rho + (\rho+p) K =0$ 
       \\
       \;\\
       \hline
       \;\\
       Scalar & $\phi \to \phi + c$ & $\frac{\rm D}{{\rm D}\tau}\rho + (\rho+p) K =0$ & \begin{tabular}{c} $\omega_{\mu\nu}=0$ \\ $(\rho+p)a_{\mu}+\mathcal{D}_{\mu}p = 0$ \end{tabular}
       \\
       \\
       \hline
  \end{tabular}
\end{table}

Let us give a hydrodynamical interpretation of the 2-form field. The equation of motion of the self-interacting Kalb-Ramond field is given by the form
\be
    \varepsilon^{\mu\nu\rho\sigma}\nabla_{\rho}\left( \sqrt{-2Y} F_Y u_{\sigma} \right) = 0 \,.
    \label{2formeom}
\ee
By projecting along $u_{\mu}$, we find
\be
    \sqrt{-2Y} F_Y \varepsilon^{\nu\rho\sigma} \nabla_{\rho}u_{\sigma} = \sqrt{-2Y} F_Y \varepsilon^{\nu\rho\sigma} \omega_{\rho\sigma} = 0 \,,
\ee
where $\varepsilon^{\nu\rho\sigma}\equiv u_{\mu}\varepsilon^{\mu\nu\rho\sigma}$ is the three-dimensional Levi-Civita tensor in the sense that $u_{\nu}\varepsilon^{\nu\rho\sigma}=0$ and $\omega_{\mu\nu}\equiv h_{[\mu|}{}^{\alpha}\nabla_{\alpha}u_{|\nu]}$ with $h_{\mu\nu}\equiv g_{\mu\nu}+u_{\mu}u_{\nu}$ is the vorticity of the congruence described by $u_{\mu}$. Therefore, a part of the equation of motion concludes
\be
    \omega_{\mu\nu} = 0 \,,
    \label{zerovorticity}
\ee
as long as $Y<0$ and $F_Y \neq 0$. The remaining part of the equation of motion is obtained by taking the Hodge dual of \eqref{2formeom} with multiplying $u^{\mu}$, yielding
\be
\frac{1}{\sqrt{-2Y}} \left[ -2YF_Y a_{\mu} + \mathcal{D}_{\mu}(F-2YF_Y) \right] =0
\,,
\label{Eulereq1}
\ee
where $a^{\mu}\equiv n^{\nu}\nabla_{\nu}u^{\mu}$ is the acceleration and $\mathcal{D}_{\mu}\equiv h^{\nu}{}_{\mu}\nabla_{\nu}$ is the derivative with respect to the direction orthogonal to the four-velocity $u^{\mu}$. The energy density $\rho$ and the pressure $p$ are read from the energy-momentum tensor \eqref{2form-T} as follows:
\be
\rho = -F\,, \quad p =F-2YF_Y
\,,
\ee
which yield 
\begin{align}
    T^{\mu\nu}=(\rho + p)u^{\mu}u^{\nu}+p g^{\mu\nu}
    \,.
\end{align}
Hence, \eqref{Eulereq1} is nothing but the general-relativistic Euler equation
\be
 (\rho+p)a_{\mu}+\mathcal{D}_{\mu}p = 0
 \,, \label{Eulereq2}
\ee
provided $Y<0$. On the other hand, within the leading-order theory $F=F(Y)$, the off-shell identity $\nabla_{\mu}J^{\mu}=0$ is equivalent to the continuity equation:
\begin{align}
    \sqrt{-2Y}F_Y \nabla_{\mu}J^{\mu}
    =\frac{\rm D}{{\rm D}\tau}\rho + (\rho+p) K =0
\end{align}
where $\frac{\rm D}{{\rm D}\tau} \equiv u^{\mu}\nabla_{\mu}$ and $K \equiv \nabla_{\mu}u^{\mu}$ are the material derivative and the expansion, respectively. In summary, the irrotational property and the Euler equation are obtained from the equation of motion while the continuity equation holds off-shell in the 2-form description, which we summarise in Table~\ref{table:fluidduality}.

As we have mentioned in \eqref{PXdp=dB}, the off-shell equation is dualised to the on-shell equation and vice versa under the duality transformation. The number density, the four-velocity, the energy density, and the pressure in the scalar description are given by
\bea
n_{\rm dual} &=& \mP_X \sqrt{2X}\,, \quad u_{\rm dual}^{\mu}= -\frac{\partial^{\mu} \phi}{\sqrt{2X} }
\,, \\
\rho_{\rm dual} &=& -\mP+2X \mP_X\,, \quad p_{\rm dual}=\mP
\,,
\eea
respectively. One can easily show that the vorticity is zero and the Euler equation trivially holds while the equation of motion of the scalar field takes the following form:
\be
\nabla_{\mu}J_{\rm dual}^{\mu}=0\,, \quad J_{\rm dual}^{\mu} \equiv n_{\rm dual} u_{\rm dual}^{\mu}
\,,
\ee
Recall that the $K$-essence is shift symmetric, concluding the existence of the conserved current $J_{\rm dual}^{\mu}$, i.e.~the current of fluid particles. The conservation $\nabla_{\mu}J^{\mu}_{\rm dual}=0$ gives the continuity equation provided that $X>0$ and $\mP_X\neq 0$.

\section{Examples}
\label{sec:examples}
In this section, we shall examine several concrete theories to see how the duality works. As we have developed in \S.~\ref{sec:dualisation}, the dual to the $K$-essence is straightforwardly obtained by the following two steps:
\begin{enumerate}
\item For a given function $\mP(X)$, we first need to solve $-\mP_X^2(X)X=Y$ for $X$ in terms of $Y$;
\item We insert the solution $X(Y)$ into $F(Y)=\mP(X(Y))-2X(Y)\mP_X(X(Y))$.
\end{enumerate}
The inverse transformation is obtained by the replacements, $X\leftrightarrow Y$ and $\mP \leftrightarrow F$. Here, the implicit function theorem guarantees the existence of the solution $X=X(Y)$ at least locally as long as \eqref{regular_con} are satisfied
and, then, under such conditions the dualisation is always possible in the sense of the field space. A cautionary remark that is worth keeping in mind is that, in general, the dualisation can permit several dual branches of solutions, although in many cases, physical requirements allow to select a physical branch. For instance, one can impose analyticity of the dualisation and require that it reduces to the identity transformation at leading order for small fields, so the dualisation can be defined perturbatively around this point.

\subsection{Power law}

We can obtain the dual of some particular theories by following the above procedure. We begin with the simplest example, the power law theory described by $\mP(X)=g_n X^n$, that corresponds to a constant equation of state $w\equiv p/\rho=P/(2XP_X-P)=1/(2n-1)$. The solution of $-\mP_X^2(X)X=Y$ is given by
\be
X=\left(-\frac{1}{g_n^2 n^2}Y\right)^{\frac{1}{2n-1}}
\ee
and, thus, the dual theory is
\be
F(Y)=g_n (1-2n)\left(-\frac{1}{g_n^2 n^2}Y\right)^{\frac{n}{2n-1}}\,.
\label{eq:powerlawdual}
\ee
We clearly notice that for $n=1$ it gives the dual relation for the free theories; that is, $P(X)=g_1 X$ is dual to $F(Y)=g_1^{-1} Y$, as already done in \S~\ref{sec:dualisation}. 
The function \eqref{eq:powerlawdual} gives the dual formulation of the symmetric superfluids obtained in \cite{Pajer:2018egx} except for the Dirac-Born-Infeld (DBI) theory that we will dualise below. 
On the other hand, the theory $n=1/2$, which corresponds to $P_X+2XP_{XX}=0$, leads to a singular dualisation. This value corresponds to the cuscuton model~\cite{Afshordi:2006ad} that in turn represents a singular theory with special properties, like the absence of the scalar degree of freedom in the unitary gauge $\phi=\phi(t)$~\cite{Gomes:2017tzd}. It has also been shown to belong to a special class of shift-symmetric scalar field theories from the symmetry perspective~\cite{Pajer:2018egx}. The theory has an (infinite) extended symmetry of the form
\be
\delta\phi=\epsilon^\mu\partial_\mu g(\phi)
\label{eq:cuscutonsym}
\ee
for an arbitrary function $g(\phi)$ and with $\epsilon^\mu$ the transformation parameter. Under this transformation, the Lagrangian of the cuscuton changes by a total derivative $\delta\Lag \propto \partial_\mu(\sqrt{X} g'\epsilon^\mu)$. The DBI theory and the cuscuton theory provide interesting insights on the scalar/2-form duality as we will discuss shortly.

\subsection{DBI}
\label{sec:DBI}

Another example is the string inspired DBI field theory, which was initially introduced as a non-linear electrodynamic model to avoid problems with divergence of the electron self-energy~\cite{Born:1934gh}. The scalar field realisation of the DBI theory is described by $\mP(X)=\Lambda^4\left(1-\sqrt{1-2X/\Lambda^4} \right)$ where $\Lambda$ is a mass scale of the theory. For this theory, the relation between the scalar field variable $X=-\frac12 \partial_\mu\phi\partial^\mu \phi$ and the Kalb-Ramond field variable  $Y=-\frac{1}{12} H_{\mu\nu\rho} H^{\mu\nu\rho}$ is 
\be
Y=-\frac{X}{1-2X/\Lambda^4} \quad \Leftrightarrow \quad  X=-\frac{Y}{1-2Y/\Lambda^4}
\,.
\label{eq:DBI_X_Y}
\ee
The dual theory $F(Y)$ to this DBI model follows from  
\be
F(Y)=\Lambda^4\left(1-\sqrt{1-2Y/\Lambda^4} \right)\,,
\ee
where we have used~\eqref{eq:DBI_X_Y}. Hence, the DBI theory is ``self-dual''; the DBI theory of the scalar field is dual to the DBI theory of the 2-form.

Let us consider extreme limits of the DBI theories to see the singular points of the dualisation. It is easy to obtain the following relations between both dual formulations:
\bea
\mP_X&=&\frac{1}{(1-2X/\Lambda^4)^{1/2}}\,, \quad \mP_X+2X\mP_{XX}=\frac{1}{(1-2X/\Lambda^4)^{3/2}}
\,, \\
F_Y&=&\frac{1}{(1-2Y/\Lambda^4)^{1/2}}\,, \quad~ F_Y+2YF_{YY}=\frac{1}{(1-2Y/\Lambda^4)^{3/2}}
\,.
\eea
The relation~\eqref{eq:DBI_X_Y} leads to the following correspondence:
\bea
2X/\Lambda^4 \to - \infty  \quad &\Leftrightarrow& \quad  2Y/\Lambda^4 \to 1
\,, \label{DBI_limit1} \\
2Y/\Lambda^4 \to - \infty  \quad &\Leftrightarrow& \quad  2X/\Lambda^4 \to 1
\,, \label{DBI_limit2}
\eea
where the limit is taken so that the inside of the square root is positive.\footnote{Recall that, in our convention, the timelike gradient of the scalar corresponds to $X>0~(\Leftrightarrow Y<0)$ and the spacelike gradient is $X<0~(\Leftrightarrow Y>0)$, respectively. Hence, the extreme limits \eqref{DBI_limit1} and \eqref{DBI_limit2} are taken under the spacelike configuration and the timelike configuration, respectively. If one wants to consider other extreme limits $2X/\Lambda^4 \to + \infty$ and $2Y/\Lambda^4 \to + \infty $, one should start from the DBI theory with a ``wrong'' sign, $\mP(X)=\Lambda^4\left(-1+\sqrt{1+2X/\Lambda^4} \right)$. } Therefore, even though the functional forms are the same, we are taking the different limits in the scalar description and the 2-form description. In particular, taking the extreme limits $2X/\Lambda^4 \to - \infty$ and $2Y/\Lambda^4 \to - \infty $ lead to the cuscuton-type Lagrangians,
\bea
\mP(X) &\to& -\Lambda^4\sqrt{-2X/\Lambda^4}
\,,  \\
F(Y) &\to & -\Lambda^4\sqrt{-2Y/\Lambda^4}
\,, 
\eea
which are located at the opposite extreme limits of the DBI theory, respectively.

\subsection{Cuscuton}

We now consider the cuscuton-type Lagrangians:
\bea
\mP(X) &=& \pm \Lambda^4 \sqrt{\pm 2X}
\,, 
\label{cus_scalar} \\
F(Y) &=& \pm \Lambda^4 \sqrt{\pm 2Y}
\,,
\label{cus_2form}
\eea
where $X$ and $Y$ are set to be dimensionless and the overall sign is chosen so that the null energy condition is respected, i.e.~$\mP_X>0$ and $F_Y>0$. The functional forms \eqref{cus_scalar} and \eqref{cus_2form} arise as solutions of $\mP_X+2X \mP_{XX}=0$ and $F_Y+2Y F_{YY}=0$, respectively, so these theories are at the singularity of the dualisation. 

Since the cuscuton theory \eqref{cus_scalar} has been studied in the literature~\cite{Afshordi:2006ad,Gomes:2017tzd}, we shall concentrate on the 2-from theory \eqref{cus_2form}, especially the cosmologically relevant case, the theory \eqref{cus_2form} with the negative sign. We have already encountered this theory: it appears as an EFT of the weak coupling limit of the $U(1)$ scalar in \S.~\ref{sec:weak/strong} and it also arises as the extreme limit of the DBI theory in \S.~\ref{sec:DBI}. As we shall argue below, the theory \eqref{cus_2form} with the negative sign is regarded as an EFT for the dust fluid, such as the dark matter component of the universe, that cannot be achieved by the shift-symmetric scalar field. 

We consider the homogeneous and isotropic background in which the field strength of the 2-form is given by $H_{ijk}=B\epsilon_{ijk}$. Hence, we obtain
\be
Y=-\frac{B^2}{2a^6(t)}
\ee
where $a(t)$ is the scale factor of the universe. The energy density, pressure and the sound speed of the theory $F=-\Lambda^4 \sqrt{-2Y}$ are
\bea
\rho &=& -F =  \frac{\Lambda^4 B}{a^3}
\,, \\
p &=& F-2Y F_Y = 0 
\,, \\
c_s^2 &=&\frac{F_Y+2YF_{YY}}{F_Y} = 0
\,,
\eea
respectively. The pressure and the sound speed vanish while the energy density decays as $a^{-3}$, which are exactly the properties of the dust fluid. In the case of the $K$-essence, the dust limit corresponds to a singular limit $\mP \to 0$. On the other hand, the dual description admits a regular expression $F=-\Lambda^4 \sqrt{-2Y}$ which can be used for the action of the irrotational dust fluid.
In realistic situations, the fluid must have small but non-zero pressure and sound speed which can be introduced by adding corrections to \eqref{cus_2form}. In fact, we have already seen that the action is corrected in the $U(1)$ example \eqref{U(1)}. As shown in \eqref{F_freeU(1)}, the finite correction of the quartic self-interaction leads to the term linear in $Y$ at the leading order which gives rise to
\be
p\propto a^{-6}
\,, \quad
c_s^2\propto a^{-3}
\,.
\ee
Furthermore, we have a higher derivative correction
\be
-\frac{F_{YY}}{2\mM^2}(\partial Y)^2 = - \frac{\Lambda^4}{2\mM^2} \frac{(\partial Y)^2}{(-2Y)^{3/2}} 
,
\ee
providing a non-relativistic dispersion relation of the fluid. From the EFT perspective, we do not need to specify the origin of the corrections (and we do not need to stick to these particular forms of the corrections). The corrections to the 2-from cuscuton universally describe deviations from the dust fluid.

\subsection{Ghost condensate}

As the final example, we consider the ghost condensate~\cite{Arkani-Hamed:2003pdi,Arkani-Hamed:2003juy} which is also a singular point of the dualisation. The ghost condensate is defined by the point $\mP_X=0$ in which the background energy-momentum tensor of the $K$-essence field behaves as a cosmological constant. The scalar field $\phi$ has a non-trivial background expectation value $X\neq 0$ as a root of $\mP_X=0$ which spontaneously breaks the Lorentz invariance. Around the background $\phi=\bar{\phi}(t)$, the quadratic action of the perturbations $\delta \phi$ (under the gravity decoupling limit) is generically given by
\begin{align}
\mS=\frac{1}{2}\int \dd t \dd k^3 \left[  (\mP_X+2X\mP_{XX}) \delta \dot{\phi}_k^2 - \mP_X k^2 \delta \phi^2_k \right],
 \label{qSk-essence}
\end{align}
 in Fourier space. The ghost condensate $\mP_X =0$ thus corresponds to the vanishing sound speed which would signal the strong coupling. Nonetheless, a small sound speed implies that a higher derivative correction, say $(\Box \phi)^2$, can not be ignored. Although the operator  $(\Box \phi)^2$ yields an additional ghostly state due the higher time derivative, the anisotropic scaling of the spacetime implies that the higher time derivative is still negligible while only the higher spatial derivative is relevant. As a result, the quadratic action of the ghost condensate is given by
\begin{align}
\mS_{\rm GC}=\frac{1}{2}\int \dd t \dd k^3 \left[ 2X\mP_{XX} \delta \dot{\phi}_k^2 - \frac{1}{\Lambda^2} k^4 \delta \phi^2_k +\cdots \right],
 \label{qSGC}
\end{align}
with the cutoff $\Lambda$ where $\cdots $ denotes more higher derivative operators. The ghost condensate provides a non-linear dispersion relation $\omega^2\propto k^4$ without the strong coupling.

Let us then consider the similar limit $F_Y \to 0$ in the 2-form theory. The background similarly behaves like a cosmological constant, while the perturbations exhibit a different property. As easily seen in \eqref{2-formqaction1} and \eqref{2-formqaction2}, the point $F_Y=0$ corresponds to a singular point that the kinetic term $\delta C$ vanishes and the action no longer depends on $\delta \vec{B}^T$. Hence, the 2-form theory is infinitely strongly coupled at $F_Y=0$. The vanishing kinetic term implies that the anisotropic scaling of the spacetime is opposite to the usual ghost condensate, meaning that the higher derivative correction can not resolve the strong coupling issue. We may not obtain a consistent theory at $F_Y = 0 $ in the 2-form description (see also~\cite{Dubovsky:2005xd}).

\section{Cosmological duality}
\label{sec:cosmologicalduality}
Having understood the general duality relation between the shift-symmetric $K$-essence and the self-interacting Kalb-Ramond field, we now specialise to the cosmological setup and explicitly show that the two descriptions provide identical cosmological solutions. We then discuss how the scalar/2-form duality connects seemingly inequivalent EFTs based on different symmetry-breaking patterns.

\subsection{Cosmology of a self-interacting Kalb-Ramond field}
\label{Sec:CosmologyKalbRammond}
In a cosmological scenario described by the FLRW metric
\be
\dd s^2=-\dd t^2+a^2\dd\vec{x}^2
\label{FLRWmetric}
\ee
the gravitational equations reduce to 
\bea
3H^2&=&\rho \label{Frieq1}\\
2\dot{H}+3H^2&=&-p \label{Frieq2}
\eea
where $\rho$ and $p$ are the energy density and pressure respectively (normalised with $8\pi G$). For the shift-symmetric scalar field cosmology, these quantities are
\bea
\rho_{\rm dual}&=&-\mP+2X\mP_X,\\
p_{\rm dual}&=&\mP.
\eea
The gravitational equations can then be combined to solve for $P$ and $P_X$ as
\bea
\mP&=&-H^2(3-2\epsilon) \label{eq:HX1}\\
X\mP_X&=&\epsilon H^2 \label{eq:HX2}
\label{incos}
\eea
where $\epsilon=-\dot{H}/H^2$ is the slow roll parameter, that we leave general for the moment. For the 2-form 
theory we have
\bea
\rho&=& -F,\\
p&=&F-2YF_Y.
\eea
We can again use the gravitational equations to solve for $F$ and $F_Y$ as
\bea
F&=&-3H^2,\label{eq:HY1}\\ 
YF_Y&=&-\epsilon H^2. \label{eq:HY2}
\eea
We can easily see how the duality is realised because the relation \eqref{dualPF} is identically satisfied
\be
\mP=F-2YF_Y=-H^2(3-2\epsilon).
\ee
Moreover, we can see that the following simple relation between the two descriptions holds
\be
X\mP_X=-YF_Y,
\label{eq:FYtoPX}
\ee
that allows to write
\be
F=\mP-2X\mP_X.
\ee
Finally, the relation between the second derivatives can be obtained to be
\be
F_Y+ 2Y F_{YY}=\frac{1}{\mP_X+2X\mP_{XX}}
\label{eq:FYYtoPXX}
\,.
\ee
All the results are consistent with our general argument in Sec.~\ref{sec:dualities}. The first slow roll parameter is given by
\be
\epsilon=-\frac{\dot{H}}{H^2}=3\frac{\dd \log F}{\dd\log Y}
\ee
that defines the inflationary regime. Furthermore, the second slow roll parameter that will determine if inflation can last long enough is given by
\be
\epsilon_2=\frac{\dd\log\epsilon}{H\dd t}=-6\left(1+\frac{Y F_{YY}}{F_Y}\right)+2\epsilon_1
\ee
that allows to write
\be
\frac{Y F_{YY}}{F_{Y}}=-1+\frac13\epsilon_1-\frac16\epsilon_2. 
\ee
Up to slow roll corrections, this expression relates the first and second derivatives of $F$. There is an obstruction however. If we compute the propagation speed given in \eqref{eq:stabilityMinkowski} we obtain:
\be
\cs^2=1+\frac{2Y F_{YY}}{F_Y}=-1+\frac23\epsilon_1-\frac13\epsilon_2
\ee
which cannot be positive in the slow-roll regime so that slow-roll inflation leads to a fatal gradient instability, in accordance with the analogous result for $\mP(X)$ cosmologies. By inserting the relations \eqref{eq:FYtoPX} and \eqref{eq:FYYtoPXX} we obtain
\be
\cs^2=\frac{\mP_X}{\mP_X+2X\mP_{XX}}
\ee
that is the propagation speed for a $\mP(X)$ theory. We can easily show that the Kalb-Ramond field gives rise to adiabatic perturbations by corroborating that the propagation speed indeed corresponds to $\delta p/\delta\rho$. If we parameterise the perturbed the metric as
\begin{align}
g_{00}=&-a^2(1+2\Phi),\\
g_{0i}=&\,a^2\partial_iS,\\
g_{ij}=&\,a^2\left[(1-2\Psi)\delta_{ij}+2\left(\partial_i\partial_j-\frac13\delta_{ij}\nabla^2\right)E\right].
\end{align}
the perturbed energy-momentum tensor for the 2-form field can be written as
\begin{align}
\delta T^0{}_0=&-F_Y\delta Y\\
\frac13\delta T^i{}_i=&-(F_Y+2YF_{YY})\delta Y\\
\delta T^0{}_i=&\frac23 YF_Y\partial_i\big(3S-B'\big)\\
\delta T^i{}_0=&-\frac23YF_Y\partial^iB'
\end{align}
where
\be
\delta Y=-\frac23 Y(9\psi+\nabla^2B)
\ee
These expressions allow to straightforwardly compute the sound speed as
\be
\cs^2=\frac{\delta p}{\delta\rho}=1+2Y\frac{F_{YY}}{F_Y}
\ee
that recuperates the result obtained in \eqref{eq:stabilityMinkowski}. We can also check that $\cs^2=\dot{p}/\dot{\rho}=\partial_Yp/\partial_Y\rho$ as it corresponds for a perfect fluid with adiabatic perturbations. 

\subsection{Duality and cosmological EFTs}

Let us discuss the cosmological spacetime with perturbations from the EFT perspective. Although it is straightforward to compute the perturbed action in both the scalar and the 2-form descriptions, we shall use the techniques developed in \cite{Cheung:2007st,Aoki:2021wew,Aoki:2022ipw}: we choose the unitary gauge where the degree(s) of freedom of the field (either the scalar or the 2-form) are eaten by the spacetime metric and express the action in terms of the metric only. The unitary gauge action can clarify how different realisations of cosmological scenarios are related.

As for the scalar field, the perturbation $\delta \phi(t,\vec{x}) \equiv \phi(t,\vec{x})-\bar{\phi}(t)$ transforms non-linearly under the time diffeomorphism, $t \to t+ \zeta^0(t,\vec{x})$, where $\bar{\phi}(t)$ is the background configuration of the scalar field. The time reparametrisation symmetry $t\to t'(t)$ allows us to set $\bar{\phi}=t$. On top of that, we can choose the gauge $\delta \phi = 0$ by the use of the freedom of $\zeta^0$ in which
\be
    X=-\frac{1}{2}g^{\mu\nu}\partial_{\mu}\phi \partial_{\nu}\phi = -\frac{1}{2}g^{00}
    \,.
\ee
Hence, the action of the $K$-essence coupled to GR in the unitary gauge is given by
\be
    \mS_{\rm dual}[g]=\int \dd^4x \sqrt{-g}\left[ \frac{1}{16\pi G}R[g] + \mP(g^{00}) \right]
    \,. \label{EFTk}
\ee
The (residual) symmetries of \eqref{EFTk} are the spatial diffeomorphisms
\be
    x^i \to x^i + \zeta^i(t,\vec{x})
    \,,
    \label{spatialdiff}
\ee
and the global time translation
\be
t \to t +c
\,, \label{timetrans}
\ee
where the latter one is originated from the diagonal part of the time translation and the shift symmetry of the scalar field $\phi \to \phi + c$ because we have identified $\phi$ with the time coordinate.

We then study the 2-form theory. The infinitesimal spacetime diffeomorphisms of the 2-form field with the parameters $\zeta^0(t,\vec{x})$ and $\zeta^i(t,\vec{x})$ around the background $\bar{B}_{ij}=\frac{1}{3}B\epsilon_{ijk}x^k$ are
\bea
\delta_\zeta B_{0i}&=&\partial_0\zeta^j \Bb_{ij}\,, \\
\delta_\zeta B_{ij}&=&-\zeta^k\partial_k\Bb_{ij}+2\partial_{[i}\zeta^k\Bb_{j]k}\,,
\eea
implying that the perturbations of $B_{\mu\nu}$ are invariant under the infinitesimal time diffeomorphism.
On the other hand, the gauge transformation with $\theta_i$ and $\theta_0$ are 
\bea
\delta_\theta B_{0i}&=&\partial_0\theta_i-\partial_i\theta_0 \,, \\
\delta_\theta B_{ij}&=&2\partial_{[i}\theta_{j]} \,.
\eea
We can choose the unitary gauge of the 2-form where the field configuration is fixed to be
\be
    B_{ij}=\frac{1}{3}B\epsilon_{ijk}x^k \,, \quad B_{0i}=0
    \,,
    \label{unitarygauge2form}
\ee
even in the presence of the perturbations. The unitary gauge yields
\be
    Y=-\frac{1}{12}g^{\mu\alpha}g^{\nu\beta}g^{\rho\gamma}H_{\mu\nu\rho}H_{\alpha\beta\gamma}
    = -\frac{B^2}{2}{\rm det}g^{ij}
    \,.
\ee
The gauge choice \eqref{unitarygauge2form} can be achieved by the freedom of $\theta_i,\theta_0$ and $\zeta^i$. Let us denote the perturbations of the 2-form before the gauge transformations by $\delta B_{0i}$ and $\delta B_{ij}$, respectively. We thus want to find a solution to
\bea
   \delta B_{0i}+\partial_0 \theta_i -\partial_i \theta_0 +\partial_0\zeta^j \Bb_{ij} &=& 0
   \,, \\
   \delta B_{ij}+2\partial_{[i}\theta_{j]} -\zeta^k\partial_k\Bb_{ij}+2\partial_{[i}\zeta^k\Bb_{j]k} &=& 0
   \,,
\eea
for arbitrary $\delta B_{0i}$ and $\delta B_{ij}$. A solution is explicitly constructed to be 
\be
 \theta_0 =0 \,, \quad
 \theta_i = -\zeta^j \Bb_{ij}-\int \dd t \, \delta B_{0i} \,,
\ee
with
\be
 \zeta^i =\frac{1}{2}\epsilon^{ijk}\left( \delta B_{jk}-2 \int \dd t\, \partial_j \delta B_{0k} \right)
 \,.
\ee
Note that the unitary gauge condition does not uniquely determine the gauge parameters $\theta_i,\theta_0$ and $\zeta^i$, meaning that there are residual gauge freedoms. The one is associated with the redundancy of the gauge symmetry of the 2-form $\theta_0 \to \theta_0+\dot{\theta}, \theta_i \to \theta_i + \partial_i \theta$ which is not important for the later discussion because this trivially acts. The important residual symmetry transformation is the following combined transformation
\be
    \zeta^i = \epsilon^{ijk}\partial_j \zeta^V_k(\vec{x}) \,, \quad
    \theta_i = -\frac{2}{3}B\left(\zeta^V_i(\vec{x}) - \frac{1}{2}x^j \partial_j \zeta^V_i(\vec{x}) \right)
    \,,
    \label{transcomb}
\ee
that preserves the unitary gauge condition \eqref{unitarygauge2form}. Here, we emphasise that $\zeta^V_k(\vec{x})$ is a function of the spatial coordinates only and $\zeta^i = \epsilon^{ijk}\partial_j \zeta^V_k(\vec{x})$ represents the infinitesimal volume-preserving diffeomorphisms. 
As a result, the unitary gauge action of the self-interacting Kalb-Ramond field is given by
\be
 \mS[g] = \int \dd^4x \sqrt{-g}\left[ \frac{1}{16\pi G}R[g] + F({\rm det}g^{ij}) \right]
    \,. \label{EFT2}
\ee
where the residual symmetries are the time diffeomorphism
\be
t\to t+\zeta^0(t,\vec{x})
\,,
\label{timediff}
\ee
and the volume-preserving diffeomorphisms
\be
x^i \to x'{}^i(\vec{x}) \quad {\rm s.t.} \quad {\rm det} \frac{\partial x'{}^i}{\partial x^j} = 1
\,.
\label{vdiff}
\ee

We have established the duality between the $K$-essence and the self-interacting Kalb-Ramond field in Sec.~\ref{sec:dualities}. In \eqref{EFTk} and \eqref{EFT2}, both actions are given by the same variable, namely the spacetime metric, and the only difference is the residual symmetries, leading to the duality of cosmology: cosmology with the residual symmetries \eqref{spatialdiff} and \eqref{timetrans} is dual to one with \eqref{timediff} and \eqref{vdiff}.

Note, however, that the unitary gauge action \eqref{EFT2} is not complete because our gauge condition is an incomplete gauge fixing. As is well-known, when one incompletely fixes the gauge at the level of action, the action cannot reproduce all of original equations (see Appendix~\ref{sec:gaugefix} and \cite{Motohashi:2016prk} for more discussions). Indeed, the action \eqref{EFT2} takes exactly the same form as the unitary gauge action of the ordinary perfect fluid: the action of the ordinary perfect fluid is given by a function of the determinant of $B^{ab}=g^{\mu\nu}\partial_{\mu}\phi^a \partial_{\nu}\phi^b$ and the unitary gauge $\phi^i=x^i$ gives $B^{ij}={\rm det} g^{ij}$. Therefore, just considering the action \eqref{EFT2}, the vorticity does not necessarily vanish. In the case of the perfect fluid, an irrotational fluid remains irrotational (Helmholtz's theorem) which is essentially a consequence of the volume-preserving diffeomorphism invariance~\cite{Dubovsky:2005xd,Aoki:2022ipw}. Hence, imposing $\omega_{\mu\nu}=0$ on \eqref{EFT2} is consistent and, in fact, $\omega_{\mu\nu}=0$ is a missing part of the equation of motion of the 2-form field.\footnote{One can recover the Euler equation \eqref{Eulereq2} that is the spatial component of the energy-momentum conservation $\nabla_{\mu}T^{\mu\nu}=0$ thanks to the Bianchi identity.} Our precise statement is that cosmology with the residual symmetries \eqref{spatialdiff} and \eqref{timetrans} is dual to one with \eqref{timediff} and \eqref{vdiff} under the condition $\omega_{\mu\nu}=0$.\footnote{In the unitary gauge, the four-velocity is given by $u^{\mu}=\delta^{\mu}_0/\sqrt{-g_{00}}$ and the vorticity is computed from the components of the metric. See~\cite{Aoki:2022ipw} for details.} The vorticity is allowed if we downgrade the gauge symmetry to the global symmetry. This also re-establishes the superfluid/fluid relation discussed in~\cite{Dubovsky:2005xd} from the point of view of the symmetry. We will revisit the precise relation between the superfluid and the fluid in Sec.~\ref{sec:superfluid}.

\section{Multi-Kalb-Ramond}
\label{sec:multiKR}

So far we have only considered one single Kalb-Ramond field. It is straightforward to extend the framework to a system with multiple Kalb-Ramond fields. In this section we will particularly focus on a set of Kalb-Ramond fields $B^a{}_{\mu\nu}$ $(a=1,2,3)$ featuring an internal global $SO(3)$ in addition to the gauge symmetry acting on each field. Thus, the physical objects will be $H^a{}_{\mu\nu\rho}=3\partial_{[ \mu}B^a{}_{\nu\rho]}$. In this framework it is possible to obtain homogeneous and isotropic solutions with the field configuration\footnote{There is of course another isotropic configuration given by $B^a{}_{ij}=\phi^a\epsilon_{ijk}x^k$  that realises the isotropy and homogeneity by only using the gauge symmetries of the 2-forms, i.e., it is nothing but several copies of the single 2-form case. Thus, we will focus on the most interesting configuration \eqref{eq:multiconfig} for our purposes here.}
\be
B^a{}_{ij}=A(t)\epsilon^a{}_{ij}\,.
\label{eq:multiconfig}
\ee
While the homogeneity is trivially realised, the isotropy is realised by a combination of the internal $SO(3)$ and the spatial rotations. In order to build the theory, we will consider the most general Lorentz- and $SO(3)$-invariant action at lowest order in derivatives. Lorentz invariance only leaves the object 
\be
M^{ab}=-\frac16H^a{}_{\mu\nu\rho} H^{b\mu\nu\rho}=\dualH^a{}_{\mu}\dualH^{b\mu}
\ee
while the internal $SO(3)$ symmetry imposes to use scalars of $M^{ab}$. Since this is a $3\times3$ matrix, we have 3 independent scalars that we parameterise as
\be
Y_1=[M],\quad Y_2=\frac{[M^2]}{[M]^2}\quad \text{and}\quad Y_3=\frac{[M^3]}{[M]^3}\,,
\label{defYiM}
\ee
and our theory will be described by the action
\be
\mS=\int\dd^4x\sqrt{-g}\,\mK(Y_1,Y_2,Y_3)\,.
\label{multiKR}
\ee
The advantage of using this parameterisation is that, in a FLRW metric and with the configuration \eqref{eq:multiconfig} we have that the only non-vanishing components of the 2-form field strength (and its dual) are
\be
H^a{}_{0ij}=\dot{A}\epsilon^a{}_{ij}\quad\Rightarrow\quad\dualH^a{}_i
=\frac{\dot{A}}{aN}\delta^a{}_i
\ee
so that the fundamental matrix reads
\be
M^{ab}=\frac{\dot{A}^2}{a^4N^2}\delta^{ab}
\label{eq:Mback}
\ee
and the introduced $SO(3)-$scalars are
\be
Y_1=3\frac{\dot{A}^2}{a^4N^2},\quad Y_2=\frac13,\quad Y_3=\frac19\,.
\ee
This means that only $Y_1$ is sensitive to the universe expansion, thus being the only relevant for the homogeneous background evolution. 
In other words, the background evolution will only depend on $\mK_{Y_1}$. This is apparent by looking at the energy-momentum tensor, which is given by
\be
T_{\mu\nu}=\mK_{ab}H^a{}_{\mu\alpha\beta} H^{b\;\;\alpha\beta}_{\;\;\nu}+\mK g_{\mu\nu}
\ee
with
\bea
\mK_{ab}&\equiv&\frac{\partial\mK}{\partial M^{ab}}=\mK_{Y_1}\delta_{ab}+\sum_{n=2}^3\frac{n\mK_{Y_n}}{[M]^{n+1}}\Big[[M](M^{n-1})_{ab}-[M^n]\delta_{ab}\Big]\\
&=&\mK_{Y_1}\delta_{ab}+\frac{2\mK_{Y_2}}{Y_1^2}\Big(M_{ab}-Y_1Y_2\delta_{ab}\Big)+\frac{3\mK_{Y_3}}{Y_1^3}\Big(M_a{}^c M_{cb}-Y_1^2Y_3\delta_{ab}\Big)\,.
\label{defKab}
\eea
From this expression we can obtain the useful relation
\be
\mK_{ab}M^{ab}=Y_1\mK_{Y_1}
\ee
that, together with the relation 
\be
H^a{}_{\mu\alpha\beta} H^{b\;\;\alpha\beta}_{\;\;\nu}=2\Big(\dualH^a{}_{\mu}\dualH^b{}_{\nu}-\dualH^{a\alpha}\dualH^b{}_{\alpha}g_{\mu\nu}\Big)
\ee
allows to write the energy-momentum tensor as
\be
T_{\mu\nu}=2\mK_{ab}\dualH^a{}_{\mu} \dualH^b{}_{\nu}+\Big(\mK -2Y_1\mK_{Y_1}\Big)g_{\mu\nu}\,.
\ee
It is then apparent that the second and third terms in \eqref{defKab} identically vanish for \eqref{eq:Mback}, thus explicitly showing that $i)$   the energy-momentum tensor is isotropic and $ii)$  only $\mK_{Y_1}$ contributes to the background evolution. The energy-density and pressure are given in this case by
\bea
\rho&=&-\mK+2Y_1\mK_{Y_1}\nonumber\\
p&=&\mK-\frac43Y_1\mK_{Y_1}
\eea
so we will have accelerated solutions provided
\be
\left\vert\frac{\partial\log\mK}{\partial\log Y_1}\right\vert\ll1\,.
\ee

From the form of the energy-momentum tensor, we can also obtain straightforwardly that scale invariance corresponds to
\be
T^\mu{}_\mu=0
\Rightarrow \mK\propto Y_1^{3/2}\,.
\ee
Because of our parameterisation \eqref{defYiM}, deviations from scale invariance are fully encoded in the dependence on $Y_1$, since $Y_2$ and $Y_3$ are clearly invariant under rescalings. Thus, the theory will be scale invariant if the Lagrangian takes the form
\be
\Lag_{\rm si}=F_1(Y_2,Y_3) Y_1^{3/2}+F_2(Y_2,Y_3)\,.
\ee

\subsection{A solid duality}

The above construction with three 2-form fields featuring an internal $SO(3)$ global symmetry is of course the dual description of the effective field theory for solids and the depicted scenario is the dual description of solid inflation~\cite{Endlich:2012pz}. The dualisation is straightforward to perform by following analogous steps as those for the single 2-form case, but we will sketch the procedure here for completeness. Let us recall, following the exposition in \cite{Endlich:2012pz}, that a solid can be described by a triplet of scalar fields $\varphi^a$, $a=1,2,3$ that are associated to the Lagrangian coordinates. The background state is then defined by $\langle\varphi^a\rangle=x^a$ and the homogeneity and isotropy are realised by further requiring an internal Euclidean group for the fields. Internal translations are achieved by imposing shift symmetry, while rotations only permit to use $SO(3)$-scalars in the Lagrangian. 

Starting from our multi Kalb-Ramond action, it should be clear that the dualisation proceeds similarly to the single 2-form case, but with three copies. We first rewrite \eqref{multiKR} in the first order formalism:
\be
\mS=\int\dd^4x\sqrt{-g}\left[-\frac12 \Pi^{a\mu\nu\rho} \partial_{[\mu} B^a{}_{\nu\rho]} + \mH(Y_1,Y_2,Y_3)\right]\,,
\ee
where $Y_i$ are the same expressions \eqref{defYiM} but for the matrix $M_\Pi^{ab}\equiv \star\Pi^a{}_{\mu}\star\Pi^{b\mu}$. The equation for the 2-form $\mathbf{B}^a$ is simply $\delta \mathbf{\Pi}^a=0$ which means we can write $\star\Pi^a=\dd\varphi^a$. We can then insert this relation back into the action to finally obtain the dual theory
\be
S_{\text{dual}}=\int\dd^4x\sqrt{-g}\mH(X_1,X_2,X_3)
\ee
where $X_i$ are nothing but the scalars $Y_i$ constructed for the matrix $M^{ab}_\varphi\equiv \partial_\mu\varphi^a \partial^\mu\varphi^b$.

\subsection{The perfect fluid limit}

The effective field theory for a perfect fluid can be obtained from the solid by imposing the absence of anisotropic stresses so fluid elements can freely move without costing any energy. This means that the perfect fluid corresponds to a situation where the symmetries are enhanced with internal volume preserving diffeomorphisms with respect to the solid. In our set-up, this means that the action can only depend on the determinant  $M\equiv\det M^{ab}$ so $\mK=\mK(M)$. For this case, we have
\be
\mK_{ab}=\mK_M M M^{-1}_{ab}
\ee
so the energy-momentum simplifies to
\bea
T_{\mu\nu}&=&\mK_M M M^{-1}_{ab}H^a{}_{\mu\alpha\beta} H^{b\;\;\alpha\beta}_{\;\;\nu}+\mK g_{\mu\nu}\nonumber\\
&=&2\mK_M M M^{-1}_{ab}\dualH^a{}_{\mu}\dualH^b{}_{\nu} +\Big(\mK-6\mK_M M\Big) g_{\mu\nu}\,.
\eea
We can introduce the 4-velocity
\be
u^{\mu}=-\frac{1}{6\sqrt{M}}\epsilon_{abc}\epsilon^{\mu\alpha\beta\gamma}\dualH^a_\alpha \dualH^b_\beta \dualH^c_\gamma
\ee
that is normalised $u^\mu u_\mu=-1$ and clearly satisfies the orthogonality relation $\dualH^d{}_\mu u^\mu=0$. In the configuration \eqref{eq:multiconfig} we have $H^a{}_i=\frac{\dot{A}}{aN}\delta^a{}_i$ so $\det M^{ab}=\left(\frac{\dot{A}}{aN}\right)^6\det g^{ab}$ and we can see that the sign in the above 4-velocity guarantees that it is future-oriented. Using that $T_{\mu\nu} u^\mu u^\nu=\rho$ and $T_{\mu\nu} g^{\mu\nu}=3p-\rho$, we can extract the energy density and pressure by contracting with $u^\mu$ and $g^{\mu\nu}$ as
\be
\rho=-\mK+6M\mK_M ,\quad p=\mK -4 M \mK_M\,.
\ee
We can interpret $\star \mathbf{H}^a$ as the frame of the fluid elements $\mathbf{e}^a$ so that $M^{ab}=e^a{}_\mu e^b{}_\nu g^{\mu\nu}$ is naturally associated to the metric of the internal fluid manifold. In fact, we can express the energy-momentum tensor in a more apparent form by using the following expression for the inverse of $M^{ab}$:
\be
2MM^{-1}_{ab}=\epsilon_{aij}\epsilon_{bmn}M^{im}M^{jn}\,,
\ee
so we have 
\begin{eqnarray}
2MM^{-1}_{ab}\dualH^a{}_\mu\dualH^b{}_\nu&=&\epsilon_{aij}\epsilon_{bmn} \dualH^i{}_\alpha\dualH^{m\alpha}\dualH^j{}_\beta\dualH^{n\beta}\dualH^a{}_\mu\dualH^b{}_\nu\nonumber\\
&=&2M(g_{\mu\nu}+u_\mu u_\nu)\nonumber\\
&=&2Mh_{\mu\nu}
\end{eqnarray}
where, in the last step we have identified the orthogonal projector. This equality shows that the orthogonal projector is indeed given by $\mathbf{h}=M^{-1}_{ab}\mathbf{e}^a\mathbf{e}^b$ as one would expect. Furthermore, we can insert it in the energy-momentum tensor to write it as
\be
T_{\mu\nu}=(-\mK+6M\mK_M)u_\mu u_\nu+(\mK-4M\mK_M)h_{\mu\nu}
\ee
which is the standard form for the energy-momentum tensor of a perfect fluid and we recognise the energy density and pressure obtained above. 
In terms of symmetries, we obtain that imposing the system to only depend on the determinant of $M^{ab}$ implies that the original $SO(3)$ symmetry is enhanced to an invariance under the larger $SL(4)$ group. The equation of state of the fluid is given by
\be
w=\frac{p}{\rho}=-\frac{1-4M\mK_M/\mK}{1-6M\mK_M/\mK}
\ee
while the sound speed of the perturbations is
\be
c_{\text{s}}^2=\frac{\dd p}{\dd \rho}=\frac{p'}{\rho'}=-\frac{3+4\frac{M\mK_{MM}}{\mK_M}}{5+6\frac{M\mK_{MM}}{\mK_M}}.
\ee
The duality with the three scalar fields is immediate. Let us consider the EFT of a perfect fluid described by
\be
\Lag=F(\det \Psi^{ab}),\quad\Psi^{ab}\equiv\partial_\mu\varphi^a\partial^\mu\varphi^b\,.
\ee
We perform the Legendre transformation
\be
\Lag=F(\det P^{ab})+\frac{\partial F}{\partial p^a{}_\mu}\Big(\partial_\mu\varphi^a-p^a{}_\mu\Big),\quad P^{ab}\equiv p^a{}_\mu p^{b\mu}\,,
\ee
so, after the field redefinition 
\be
\pi^{a\mu}\equiv \frac{\partial F}{\partial p^a{}_\mu}=2F'\det P\; P^{-1}_{mn}\delta^{a(m}p^{n)\mu}
\ee
so that
\be
\frac{\partial F}{\partial p^a{}_\mu}p^a{}_\mu=6F'\det P^{ab}
\ee
we obtain
\be
\Lag=\pi^{a\mu}\partial_\mu\varphi^a+\mK(\det \Pi^{ab}),\quad \Pi^{ab}\equiv \pi^a{}_\mu \pi^{b\mu}\,,
\ee
with $\mK=F-6F'\det P^{ab}$. Once again, the equation for $\varphi^a$ imposes $\mathbf{\pi}^{a}=\star\dd\mathbf{B}^a$ for some 2-form $\mathbf{B}^a$, thus showing the duality of the two descriptions for the fluid.

\subsection{Back to the superfluid}
\label{sec:superfluid}

\begin{figure}
\centering
\includegraphics[width=.8\linewidth]{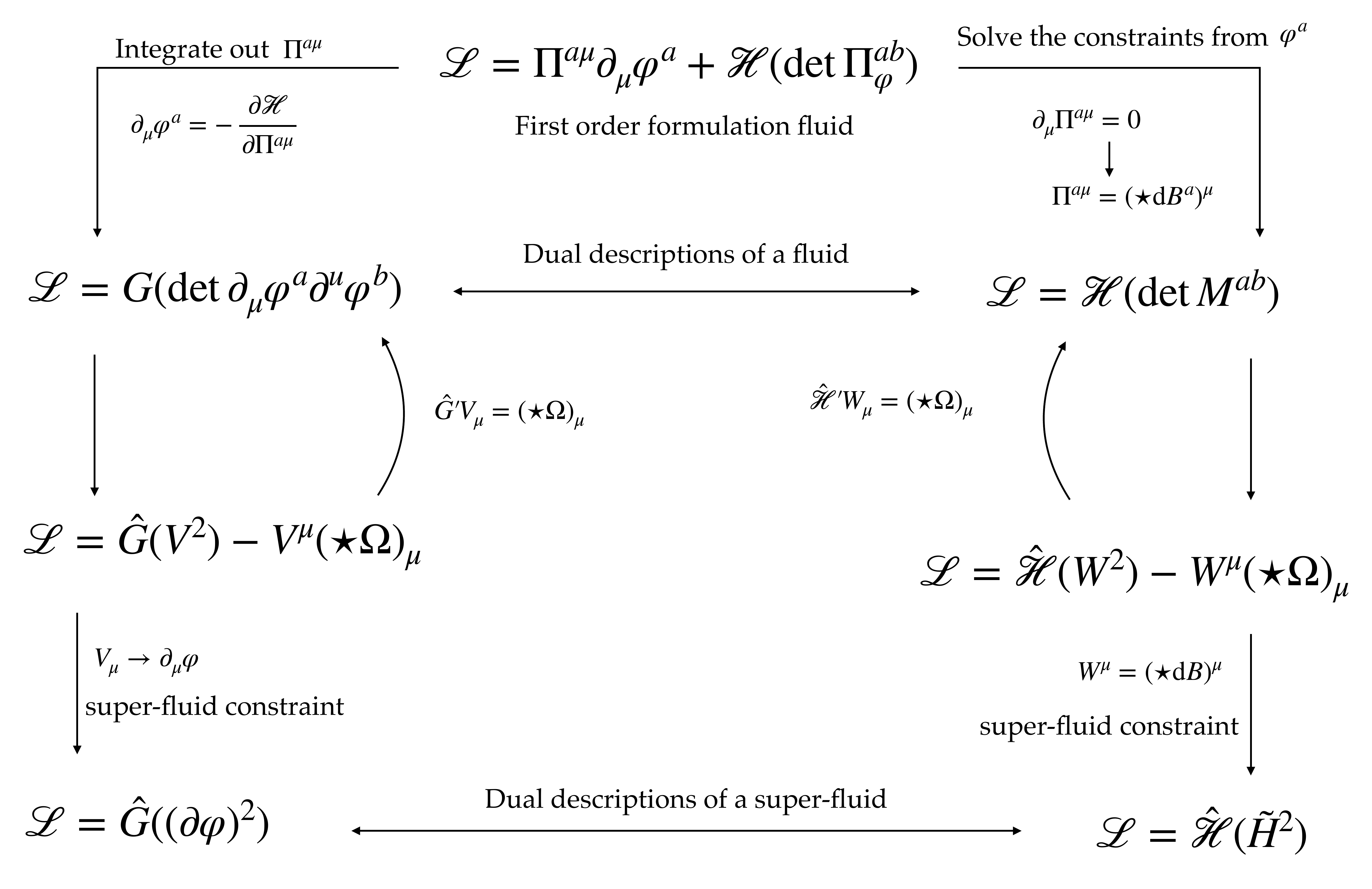}
\caption{This figure summarises the duality relations between the different formulations of a fluid and a superfluid.}
\label{Fig:Dualfluids}
\end{figure}

In the limit of vanishing vorticity, the perfect fluid reduces to the superfluid. Thus, it should be possible to obtain the EFT of a superfluid from the general EFT of perfect fluids by imposing the vanishing vorticity constraint. In this subsection we will discuss how to perform this reduction in the two dual formulations of the perfect fluid. Let us start with the usual description of the perfect fluid described with the three 2-form fields $\mathbf{B}^a$ and let us rewrite the Lagrangian as follows
\be
\Lag=\mH(V^2/2)-V^{\mu\nu\rho}\Omega_{\mu\nu\rho}
\label{eq:LagPFauxH}
\ee
with $V^{\mu\nu\rho}$ a non-dynamical 3-form field and $\Omega=\frac16\epsilon_{abc}\tilde{H}^a\wedge\tilde{H}^b\wedge\tilde{H}^c$ the volume form. The equation for $V^{\mu\nu\rho}$ reads
\be
\mH'V_{\mu\nu\rho}=\Omega_{\mu\nu\rho}\,.
\ee
Noticing that $\Omega^2=(\det M^{ab})^2$, we can use the above equation to obtain $V^2=V^2(\det M^{ab})$ so we can integrate $V^{\mu\nu\rho}$ out and obtain
\be
\Lag=\Big[\mH(V^2/2)-\mH'V^2\Big]_{V^2=V^2(\det M^{ab})}\equiv \mK(M)
\label{eq:LagPFauxH2}
\ee
so we recover the description of the perfect fluid in terms of the 2-forms. The Lagrangian \eqref{eq:LagPFauxH} is useful however because it allows to make more direct contact with the superfluid. We can add the vanishing vorticity constraint by means of a Lagrangian multiplier as follows:
\be
\Lag=\mH(V^2/2)-V^{\mu\nu\rho}\Omega_{\mu\nu\rho}+\partial_{[\mu}\lambda_{\nu\rho]}\Omega^{\mu\nu\rho}\,.
\label{eq:LagPFauxconstrainedH}
\ee
The constraint imposes that $\Omega$ has vanishing co-differential, so it can be written as $\Omega_{\mu\nu\rho}=\epsilon_{\mu\nu\rho\alpha}\partial^\alpha\lambda$ and the Lagrangian reads
\be
\Lag=\mH(V^2/2)-V^{\mu\nu\rho}\epsilon_{\mu\nu\rho\alpha}\partial^\alpha\lambda.
\label{eq:LagPFauxconstrainedH2}
\ee
The equation for $\lambda$ tells us that we can write $V_{\mu\nu\rho}=\partial_{[\mu}B_{\nu\rho]}$ so the original Lagrangian supplemented by the constraint $\delta\Omega=0$ reduces to
\be
\Lag=\mH(H^2(B))\,,
\ee
i.e., the superfluid described in terms of the 2-form field. We can proceed analogously for the dual formulation in terms of scalar fields starting with the Lagrangian
\be
\Lag=\mH(V^2/2)-V^\mu\tilde{\Omega}_\mu
\label{eq:LagPFaux}
\ee
where $V^\mu$ is a non-dynamical vector field and $\tilde{\Omega}_\mu=\epsilon_\mu{}^{\nu\rho\lambda}\Omega_{\nu\rho\lambda}$ is the dual of the volume form. Upon variation w.r.t. the auxiliary variable $V^\mu$ we obtain
\be
\mH'V_\mu=\tilde{\Omega}_\mu\,,
\ee
that allows to obtain $V^2=V^2(\det \Psi^{ab})$ so the Lagrangian can be expressed as
\be
\Lag=\Big(\mH-\mH'V^2\Big)_{V^2=V^2(\det \Psi^{ab})}\equiv\mK(\det \Psi^{ab})
\ee
and we recover the usual description of the perfect fluid. We can modified the Lagrangian with an appropriate constraint:
\be
\Lag=\mH(V^2/2)-V^\mu\tilde{\Omega}_\mu+\partial_\mu\lambda\tilde{\Omega}^\mu.
\label{eq:LagPFauxconstrained}
\ee
The constraint imposed by the Lagrange multiplier can be written as $\dd\Omega=0$ so we have $\Omega=\dd \mathbf{w}$ with $\mathbf{w}$ a 2-form. We can insert this relation into the Lagrangian to obtain
\be
\Lag=\mH(V^2/2)+\partial^{[\alpha}V^{\mu]}w_{[\mu\alpha]}\,.
\label{eq:LagPFauxconstrained2}
\ee
The equation for $\mathbf{w}$ now imposes $\mathbf{V}$ to be a closed form so it can be expressed as the exact form $\mathbf{V}=\dd\mathbf{\varphi}$. We can insert this expression in the Lagrangian and we finally obtain a theory for the single scalar field $\varphi$ that describes a superfluid. We summarise all the discussed formulations of the fluid and the superfluid in Fig. \ref{Fig:Dualfluids}.

\section{Massive Kalb-Ramond}
\label{sec:massive2form}
Although the main focus of this work is to explore the cosmologies and dualities of massless 2-form fields, we will briefly incur into the massive case in order to illustrate some crucial differences and important properties of breaking the gauge symmetry and how this translates into the dual theories. We will defer a more detailed study of the massive case for future work. 

\subsection{Duality to Proca fields}
We will commence our tour on the massive Kalb-Ramond field by considering a breaking of the gauge symmetry only on potential terms, i.e., all derivatives of the 2-form will still enter through its field strength. In this case, we can start from a first order formulation given by
\be
\mS=\int\dd^4x \sqrt{-g} \left[-\frac12 \Pi^{\mu\nu\rho}\partial_{[\mu}B_{\nu\rho]} + \mH(Y_{\Pi},B)\right]
\,, \quad B\equiv -\frac{1}{4}B_{\mu\nu}B^{\mu\nu}
\label{action_m2form}
\ee
where the explicit dependence on the 2-form in $\mH$ breaks the corresponding gauge symmetry. This prevents to solve for the momentum $\Pi$ in terms of a scalar field, since the 2-form equation of motion is now
\be
\nabla_\mu \Pi^{\mu\nu\rho}=\frac{\partial \mH}{\partial B_{\nu\rho}}.
\label{eq:massivedual}
\ee
We can still proceed with a dualisation by directly dualising the 2-form and the conjugate momentum as\footnote{This dualisation can be interpreted as a canonical transformation where the coordinates and the momenta are exchanged.}
\be
\Pi^{\mu\nu\rho}\equiv \epsilon^{\mu\nu\rho\sigma} A_\sigma,\quad\quad B_{\mu\nu}\equiv \frac{1}{2}\epsilon_{\mu\nu\rho\sigma} \Pi^{\rho\sigma}.
\ee
By inserting these field redefinitions into the action we obtain
\be
\mS=\int\dd^4x \sqrt{-g} \left[ -\Pi^{\mu\nu}\partial_{[\mu}A_{\nu]} + \mH(A,Z_{\Pi})\right]
\,, \quad
A\equiv -\frac{1}{2}A_{\mu}A^{\mu} \,, ~ Z_{\Pi} \equiv -\frac{1}{4} \Pi_{\mu\nu}\Pi^{\mu\nu}
\label{action_mvec}
\ee
which is nothing but the first order formulation of a self-interacting vector field. Here, we note that
\be
Y_{\Pi}= - A
\,, \quad
B = -Z_{\Pi}
\,,
\label{YZrel}
\ee
under the dualisation. The inverse dualisation is obtained by following the inverse path and we will not give the explicit construction here. The duality to a Proca field makes it manifest that the massive 2-form field propagates 3 degrees of freedom, i.e., two more than its massless counterpart. Given this difference in the number of degrees of freedom between the massive and massless theories, it is instructive to analyse the massless limit to see where the different degrees of freedom go.

\subsection{Massless limit}

The 2-form is dual to the scalar field, not the vector field, in the absence of the potential, implying that there must be a condition on the potential to have the 2-form/vector duality. Although \eqref{action_m2form} and \eqref{action_mvec} are equivalent, we need to check regularity conditions of the Legendre transformations. The variation of \eqref{action_m2form} with respect to $\Pi^{\mu\nu\rho}$ gives
\be
H_{\mu\nu\rho}+ \mH_{Y_{\Pi} }(Y_{\Pi},B) \Pi_{\mu\nu\rho} =0
\ee
where the subscript represents the derivative, e.g.,~$\mH_{Y_{\Pi} }  = \partial \mH/\partial Y_{\Pi}$. Squaring it yields the equation $Y=\mH_{Y_{\Pi}}^2 Y_{\Pi}$ which can be solved by $Y_{\Pi}=Y_{\Pi}(Y,B)$ under
\be
\mH_{Y_{\Pi}}( \mH_{Y_{\Pi}} + 2Y_{\Pi} \mH_{Y_{\Pi}Y_{\Pi}} )\neq 0
\,.
\label{reg_m2form}
\ee
Then, we find the solution $\Pi_{\mu\nu\rho}=-\mH_{Y_{\Pi}}^{-1}(Y_{\Pi}(Y,B),B)H_{\mu\nu\rho}$ and arrive at the second order formulation of the massive Kalb-Ramond field
\be
\mS=\int \dd^4 x \sqrt{-g} F(Y,B) \,, \quad F \equiv  \mH - 2Y_{\Pi}\mH_{Y_{\Pi}} \big|_{Y_{\Pi}=Y_{\Pi}(Y,B)} 
\,,
\ee
with the relations
\be
F_Y = -\frac{1}{\mH_{Y_{\Pi}}} \,, \quad F_Y+2YF_{YY}=-\frac{1}{\mH_{Y_{\Pi}} + 2Y_{\Pi} \mH_{Y_{\Pi}Y_{\Pi}}}
\,.
\ee
The regularity condition of the inverse transformation is given by the finiteness of the left-hand-side of \eqref{reg_m2form} or
\be
F_Y(F_Y+2YF_{YY}) \neq 0\,.
\label{reg_m2form2}
\ee
The conditions \eqref{reg_m2form} and \eqref{reg_m2form2} are present in the massless case as well while there are additional conditions from the regularity of the vector side in the massive case. The variation of \eqref{action_mvec} with respect to $\Pi_{\mu\nu}$ yields
\be
F_{\mu\nu} + \mH_{Z_{\Pi}} (A,Z_{\Pi}) \Pi_{\mu\nu} = 0\,, \quad F_{\mu\nu}\equiv 2\partial_{[\mu}A_{\nu]}
\,,
\ee
which can be solved by $\Pi_{\mu\nu}=-\mH_{Z_{\Pi}}^{-1}(A,Z_{\Pi}(A,Z) ) F_{\mu\nu}$ with $Z \equiv -\frac{1}{4}F_{\mu\nu}F^{\mu\nu}$ provided 
\be
\mH_{Z_{\Pi}}( \mH_{Z_{\Pi}} + 2Z_{\Pi} \mH_{Z_{\Pi}Z_{\Pi}} )\neq 0
\,.
\label{reg_mvec}
\ee
The second order form of the action is then
\be
\mS = \int \dd^4 x \sqrt{-g} G(Z,A)\,, \quad G(Z,A) \equiv \mH - 2Z_{\Pi} \mH_{Z_{\Pi}} \big|_{Z_{\Pi}=Z_{\Pi}(A,Z) }
\,,
\ee
where we should impose $G_Z(G_Z+2ZG_{ZZ}) = -\mH_{Z_{\Pi}}^{-1} ( \mH_{Z_{\Pi}} + 2Z_{\Pi} \mH_{Z_{\Pi}Z_{\Pi}} )^{-1} \neq 0$ for the regularity of the inverse transformation.
Since there are the relations \eqref{YZrel}, the regularity condition \eqref{reg_mvec} (or \eqref{reg_m2form}) can be written in terms of the derivatives of $\mH$ with respect to $B$ (or $A$):
\bea
\mH_B (\mH_B + 2B \mH_{BB}) &\neq& 0 \,, \\
\mH_A (\mH_A + 2A \mH_{AA}) &\neq& 0 \,.
\eea
Therefore, the regularity of the Legendre transformation in the one side yields the condition on the potential in the dual side, as expected.

Let us see the massless limit of the Kalb-Ramond field and its dual description. For simplicity, we consider the free theory:
\be
\mS = \int \dd^4 x \sqrt{-g} \left[ -\frac{1}{12}H_{\mu\nu\rho}H^{\mu\nu\rho} - \frac{m^2}{4}B_{\mu\nu}B^{\mu\nu} \right]
\,,
\label{eq:AppMassiveB}
\ee
whose dual description is
\be
\mS_{\rm dual} = \int \dd^4 x \sqrt{-g} \left[ -\frac{1}{4m^2}F_{\mu\nu}F^{\mu\nu} - \frac{1}{2}A_{\mu}A^{\mu} \right]
\label{eq:AppMassiveA}
\ee
with the on-shell relations
\be
H^{\mu\nu\rho} = \varepsilon^{\mu\nu\rho\sigma}A_{\sigma} \,, \quad B_{\mu\nu} = \frac{1}{2m^2}\varepsilon_{\mu\nu\rho\sigma}F^{\rho\sigma}
\,.
\ee
The massless limit of this theory must be treated with some care because we would encounter a discontinuity in the number of propagating degrees of freedom if taken directly from \eqref{eq:AppMassiveB} at face value. To avoid this difficulty, we will resort to the St\"uckelberg trick to first restore the gauge symmetry by making the replacement
\be
B_{\mu\nu} \to B_{\mu\nu} + \frac{2}{m}\partial_{[\mu}\mathcal{A}_{\nu]}
\ee
where $\mathcal{A}_{\mu}$ is the St\"{u}eckelberg field, and then take the limit $m\to 0$. Notice that we have actually introduced two gauge symmetries, namely: the desired one generated by a 1-form $\theta$ as $\mathbf{B}\to\mathbf{B}+\dd\mathbf{\theta}$ together with $\mathcal{A}\to\mathcal{A}-m\,\mathbf{\theta}$ and a secondary one generated by a 0-from $\vartheta$ that only affects the St\"uckelberg field $\mathcal{A}\to\mathcal{A}+\dd\mathbf{\vartheta}$. While the former corresponds to the St\"uckelberg trick, the latter stems from the gauge invariance possessed by the gauge transformation of the 2-form.

The resulting action in the decoupling limit with $m\to0$ consists of the massless Kalb-Ramond field plus a $U(1)$ gauge field described by
\be
\mS_{m\to 0}= \int \dd^4 x \sqrt{-g} \left[ -\frac{1}{12}H_{\mu\nu\rho}H^{\mu\nu\rho} - \frac{1}{4}\mathcal{F}_{\mu\nu}\mathcal{F}^{\mu\nu} \right]
\,, \quad \mathcal{F}_{\mu\nu} \equiv  2 \partial_{[\mu}\mathcal{A}_{\nu]}
\,,
\label{masslesslimit2form}
\ee
where it is apparent that the two sectors completely decouple and we maintain the correct number of propagating dof's. Here, we notice that the redundancy of the gauge transformation of the 2-form is recast as the gauge symmetry of the St\"{u}eckelberg field $\mathcal{A}_{\mu}$, while the restored gauge symmetry generated by $\theta$ only affects the 2-form so the St\"uckelberg field becomes gauge invariant in the decoupling limit, as it is apparent from its transformation $\mathcal{A}\to\mathcal{A}-m\,\mathbf{\theta}$.

We can proceed similarly with the dual formulation of the theory given by \eqref{eq:AppMassiveA} where the St\"uckelberg trick now amounts to replacing
\be
A_{\mu} \to m \mathcal{A}_{\mu} + \partial_{\mu} \phi
\,,
\ee
with $\phi$ the St\"uckelberg field, so the massless limit of the dual description gives
\be
\mS_{{\rm dual},m\to0} = \int \dd^4 x \sqrt{-g}\left[ - \frac{1}{4}\mathcal{F}_{\mu\nu}\mathcal{F}^{\mu\nu} -\frac{1}{2}\partial_{\mu}\phi \partial^{\mu}\phi \right]
\,.
\label{masslesslimitmvec}
\ee
The dual actions in the massless limit \eqref{masslesslimit2form} and \eqref{masslesslimitmvec} clearly describe the same dynamics with a gauge vector field and a scalar dof that is described by $\phi$ and the massless 2-form $B_{\mu\mu}$ respectively, thus preserving the duality relation in the massless limit. We have obtained this massless limit for the free theories, but the same can be shown for interacting theories, although in that case the coupling constants must also be appropriately included in a suitable decoupling limit. To see how this is achieved it is convenient to start from the first order formulation of the theory, so let us start from the interacting theory
\be
\mS=\int\dd^4x \sqrt{-g} \left[-\frac12 \Pi^{\mu\nu\rho}\partial_{[\mu}B_{\nu\rho]} +\frac{1}{12}\Pi_{\mu\nu\rho}\Pi^{\mu\nu\rho}-\frac14m^2B_{\mu\nu}B^{\mu\nu}+g_4(B_{\mu\nu}B^{\mu\nu})^2\right]
\,,
\label{action_g4form}
\ee
with $g_4$ some coupling constant. We can St\"uckelbergsize this action as before and take the decoupling limit 
\be
m\to0,\quad g_4\to0\quad\text{with}\quad \Lambda\equiv\frac{m}{g_4^{1/4}}\quad\text{fixed}
\ee
so the action reads
\be
\mS_{\text{dec}}=\int\dd^4x \sqrt{-g} \left[-\frac12 \Pi^{\mu\nu\rho}\partial_{[\mu}B_{\nu\rho]}+\frac{1}{12}\Pi_{\mu\nu\rho}\Pi^{\mu\nu\rho} - \frac14\mathcal{F}_{\mu\nu}\mathcal{F}^{\mu\nu}+\frac{1}{\Lambda^4}(\mathcal{F}_{\mu\nu}\mathcal{F}^{\mu\nu})^2\right]
\,,
\label{action_g4formdec}
\ee
that describes a free massless 2-from field in the first order formulation plus an interacting $U(1)$ gauge field in its second order formulation so \eqref{action_g4formdec} is a Routhian of the system. We can proceed similarly for the dual of \eqref{action_g4form} that is given by
\be
\mS_{\text{dual}}=\int\dd^4x \sqrt{-g} \left[- \Pi^{\mu\nu}\partial_{[\mu}A_{\nu]}-\frac12A_\mu A^\mu +\frac{m^2}{4}\Pi_{\mu\nu}\Pi^{\mu\nu}+g_4(\Pi_{\mu\nu}\Pi^{\mu\nu})^2\right]
\,.
\label{action_g4dual}
\ee
We can now St\"uckelbergise this action and take the same decoupling limit $m,g_4\to0$ with $\Lambda$ fixed. Before that, we will express the action in the second order formalism first. For that, we use the equation for $\Pi^{\mu\nu}$
\be
2\partial_{[\mu} A_{\nu]}=m^2\Pi_{\mu\nu}\left(1+\frac{4m^2}{\Lambda^4}\Pi^2\right) 
\ee
that, at leading order in the decoupling limit, yields:
\be
\mathcal{F}_{\mu\nu}=2\partial_{[\mu} A_{\nu]}\simeq m^2\Pi_{\mu\nu}.
\ee
We can insert this solution into the action to obtain the decoupling limit
\be
\mS_{\text{dual,dec}}=\int\dd^4x \sqrt{-g} \left[- \frac14\mathcal{F}_{\mu\nu}\mathcal{F}^{\mu\nu}+\frac{1}{\Lambda^4}(\mathcal{F}_{\mu\nu}\mathcal{F}^{\mu\nu})^2-\frac12 (\partial \phi)^2\right]
\,,
\label{action_g4dualdec}
\ee
that gives the same theory \eqref{action_g4formdec} with a self-interacting gauge vector field and the scalar field accounts for the decoupled massless 2-form in \eqref{action_g4formdec}. Thus, we see how the duality persists when appropriately taking the massless limit and, in particular, no conundrum arises due to the different number of propagating dof's for a massive and an exactly massless 2-form field.

\subsection{Cosmological configurations}

Since the 2-form is massive, we do not have internal symmetries at our disposal to construct homogeneous and isotropic solutions mixing external and internal generators. Thus, we can only use the usual realisations of a cosmological residual $ISO(3)$ in terms of the Lorentz generators. Given the nature of the massive  2-form field, it is not possible to construct homogeneous and isotropic configurations or, more precisely, the only possibility is having a trivial 2-form field $B_{\mu\nu}=0$. Any other configuration with a non-trivial $B_{0i}$ or $B_{ij}$ will necessarily introduce a preferred direction. This property might seem in contradiction with the dual formulation of the theory in terms of a Proca field. In this case, we can construct a homogeneous and isotropic configuration given by  $A_{\mu}=(A_0(t),0)$ with $A_0 \neq 0$. This apparent contradiction can be solved by looking at the regularity condition of the dualisation. The background equation of motion of the vector field is given by
\be
A_0 \mH_A = 0
\,,
\ee
where $\mH_A$ is evaluated at the background, $A_{\mu}=(A_0,0), \Pi^{\mu\nu}=0$. The branch $A_0=0$ corresponds to the trivial configuration of the 2-form $B_{\mu\nu}=0$ while the vector description possesses a different branch $A_0 \neq 0$ given by a root of $\mH_A =0$. Hence, the non-trivial cosmological solution of the vector field is the singular point of the dualisation and thus there is no corresponding solution in the 2-form description.

There is an interesting class of cosmologies based on rapidly oscillating massive vector fields. It was shown in \cite{Cembranos:2012kk} (see also the generalisations \cite{Cembranos:2012ng,Cembranos:2013cba}) that a spacelike vector field that oscillates rapidly gives rise to an energy-momentum tensor that is isotropic when averaged over several oscillations. This opens the possibility of having isotropic (averaged) cosmologies driven by massive spin 1 fields in spacelike configurations. Since a massive vector field is dual to a massive Kalb-Ramond field as long as the regularity conditions hold, it is then also possible to have cosmologies with oscillating massive Kalb-Ramond fields. It would be interesting to work out the description of these cosmologies in terms of the dual 2-form. 

\subsection{Massive multi-Kalb-Ramond}

At this point, it should be obvious to see that the massive version of the multi-Kalb Ramond will be dual to a multi-Proca field theory. We do not intend to develop the entire dualisation procedure here, but we will simply point out some interesting phenomenological consequences that the massive multi-Kalb-Ramond theories bring about and that are not present in the massless case. Let us consider a theory for three massive multi-Kalb-Ramond fields $B^{a}{}_{\mu\nu}$ with an internal SO(3) global symmetry. Then, we can consider a background configuration of the form
\be
B^a{}_{ij}=B(t) \epsilon^a{}_{ij}
\ee
that preserves a diagonal $SO(3)$ mixing internal and spatial rotations. On this background, we can perform perturbations of the 2-form fields $\delta B^a_{\mu\nu}$ and we decompose them into irreps of the unbroken rotational symmetry of the background. The important feature to notice is that we can construct the perturbation $t_{ij}\equiv\delta B^{(a}_{kl}\epsilon^{b)kl}\delta_{ai}\delta_{bj}$, whose transverse traceless part precisely transforms as a helicity-2 mode. Thus, these theories will give rise to a second helicity-2 mode that will naturally mix with the usual gravitational waves at linear order. In these scenarios, we might expect to obtain signatures in the signals of gravitational waves emitted by binary sources since they will mix with this second cosmological tensor mode in their propagation from the source to the detectors, thus giving rise to oscillations. These oscillations will imprint effects like those studied in e.g. \cite{BeltranJimenez:2019xxx,Ezquiaga:2021ler}. The presence of this second tensor mode for the considered background configuration is reminiscent of the dual version of these scenarios in terms of massive vector fields (see e.g. \cite{Caldwell:2016sut,BeltranJimenez:2018ymu,Caldwell:2018feo}). These theories support cosmological solutions with the configuration $A^a_\mu=A(t)\delta^a_\mu$. The perturbations around this background can be arranged to construct $f_{ij}\equiv\delta^a{}_{(i} \delta A^a{}_{j)}$, whose transverse traceless part gives rise to a second tensor mode that represents the dual of $t_{ij}$.

\subsection{On non-minimal couplings}

We will finalise our brief discussion of the massive 2-form theories by succinctly commenting on the possibility of constructing non-trivial and ghost-free non-minimal couplings. We should stress that we do not intend to provide an exhaustive exploration of all the possible ghost-free interactions, but we will content ourselves with outlining several approaches to obtained such interactions. Firstly, we will work with the St\"uckelberg fields and in the decoupling limit where the 2-form field essentially reduces to the field strength of the S\"ueckelbergs $B_{\mu\nu}\to \partial_{[\mu}\mathcal{A}_{\nu]}$. In this decoupling limit, it is clear that the problem reduces to obtaining non-minimal couplings for an Abelian spin-1 gauge field. Horndeski already tackled this problem in \cite{Horndeski:1976gi} where he obtained that the only allowed non-minimal term in the Lagrangian (under some assumptions) is a coupling to the double dual Riemann tensor. In our context, this implies that the 2-form field admits the ghost-free non-minimal coupling
\be
\Lag_{\text{non-minimal}}\supset L^{\mu\nu\alpha\beta} B_{\mu\nu}B_{\alpha\beta} 
\ee
with $L^{\mu\nu\alpha\beta}\equiv\frac14 \epsilon^{\mu\nu\rho\sigma}\epsilon^{\alpha\beta\gamma\delta} R_{\rho\sigma\gamma\delta}$ the double dual Riemann tensor. It is immediate to obtain that this term reduces to the Horndeski vector-tensor interaction for the St\"uckelberg fields in the decoupling limit. It is interesting to obtain the dual of this interaction because it generates a highly non-trivial non-minimal coupling for the vector field. We can write down the first order formulation of a theory for a massive 2-form field including this non-minimal coupling as
\be
\mS=\int\dd^4x \sqrt{-g} \left[-\frac12 \Pi^{\mu\nu\rho}\partial_{[\mu}B_{\nu\rho]} + \mH(Y_{\Pi},B)+a\,L^{\mu\nu\alpha\beta} B_{\mu\nu}B_{\alpha\beta}\right]\,
\label{action_n2nonminimal}
\ee
with $a$ some parameter. Since the non-minimal coupling does not depend on the conjugate momentum, we can integrate it out as in the absence of the non-minimal coupling. Instead, by performing the dualisation as defined in \eqref{eq:massivedual}, we obtain the dual action
\be
\mS=\int\dd^4x \sqrt{-g} \left[ -\Pi^{\mu\nu}\partial_{[\mu}A_{\nu]} + \mH(A,Z_{\Pi})+a\,R^{\mu\nu\alpha\beta} \Pi_{\mu\nu}\Pi_{\alpha\beta}\right]\,,
\label{action_mvecnonminimaldual2}
\ee
where now the non-minimal coupling does depend on the conjugate momentum. To obtain the second order form of the theory, we need to integrate $\Pi^{\mu\nu}$ out, whose equation is now given by
\be
F_{\mu\nu}=\Big(-\mH_{Z_{\Pi}}g_{\mu\alpha}g_{\nu\beta}+4a R_{\mu\nu\alpha\beta}\Big)\Pi^{\alpha\beta}.
\ee
This algebraic equation can be solved for $\Pi^{\mu\nu}$ and plugged back into \eqref{action_mvecnonminimaldual2} to obtain the second order formulation of the theory. However, the resulting Lagrangian will now include the non-minimal coupling in a highly non-trivial way. Since, as we have argued, the original formulation in terms of the 2-form relates to the ghost-free Horndeski interaction in the decoupling limit, the same property will persist in the dual formulation, although in a substantially less trivial manner.

Another approach to construct non-minimal couplings for the 2-form is to proceed in the reverse order, i.e., we can start from the healthy non-minimal couplings of a 1-form and then proceed to perform the dualisation to the 2-form. However, this procedure is not easy (and sometimes not even possible to do) for general non-minimal couplings. The main difficulty arises from derivative interactions that are not constructed in terms of the field strength. There is nevertheless a sub-class of interactions that can be straightforwardly dualised. To exemplify this, let us consider the non-minimal coupling $G^{\mu\nu} \mA_\mu\mA_\nu$.\footnote{A similar procedure has been used in \cite{Yoshida:2019dxu} (see also \cite{Takahashi:2019vax}) to obtain non-minimal couplings for a massless 2-form field starting from a derivative coupling of a scalar field to the Einstein tensor, which would correspond to the decoupling limit of the interaction we have considered here.} This interaction can be treated in an analogous way to the vector-Horndeski interaction. We can write down the first order formulation of the 1-form field and simply add this non-minimal coupling, which will not affect the procedure of integrating out the conjugate momentum. We can then dualise the theory to the 2-form description and, in that formulation, integrate out the corresponding conjugate momentum. Since now the dualisation of the non-minimal coupling $G^{\mu\nu} \mA_\mu\mA_\nu$ generates a term $G^{\mu\nu}\epsilon_{\mu\alpha\beta\gamma}\epsilon_{\nu\lambda\rho\sigma} \Pi^{\alpha\beta\gamma} \Pi^{\lambda\rho\sigma}$, integrating out $\Pi^{\mu\nu\rho}$ to obtain the second order formulation of the theory will generate a highly non-trivial non-minimal coupling. 

Let us finally mention that one could also use the findings of \cite{Bettoni:2015wla} for non-minimally coupled perfect fluids to construct other classes of non-minimal couplings for a massless Kalb-Ramond field since by exploiting the fact that perfect fluids can be described in terms of these fields.

\section{Discussion}
\label{sec:discussion}

In this work we have considered cosmologies driven by Kalb-Ramond fields where the homogeneity and isotropy of the universe are achieved by a combination of the usual spatial transformations and internal symmetries of the Kalb-Ramond fields. We have commenced with a single massless Kalb-Ramond field and extensively exploited its duality relation with shift symmetric scalar fields, emphasising the dual realisations of the cosmological principle in the two dual descriptions of the theories. In particular, we have discussed the usual weak/strong regimes of theories related by a duality transformation. We have also shown that slow roll inflationary solutions supported by a massless 2-form field comes hand in hand with Laplacian instabilities, thus recovering the known result for the dual scenario with shift-symmetric scalar fields. An interesting outcome of our study is the dual description of super-fluids in terms of a massless 2-form and the description of the EFT of cosmological perturbations in terms of the 2-form. We have worked out some explicit examples of dualities and pointed out the singular character of some theories like the cuscuton for which the dualisation procedure is ill-defined. This model belongs to the class of symmetric superfluids analysed in \cite{Pajer:2018egx}. It would be interesting to obtain the duals of the generic symmetric superfluids unveiled in that work and understand the dual realisation of the symmetries. 

We have then studied a scenario with multipole Kalb-Ramond fields with an internal $SO(3)$ symmetry. In this context, we have seen how the internal $SO(3)$ symmetry permits to construct another class of cosmological solutions with a non-trivially realised cosmological principle. We have explicitly constructed the dualised theory in terms of three scalar fields that make it apparent how the considered multi-Kalb-Ramond theory represents the dual realisation of the EFT of a solid. Within this scenario, we have analysed the perfect fluid limit with an enhanced symmetry in the two dual formulations with scalar and 2-form fields. Finally, the more symmetric super-fluid limit has also been explicitly constructed, recovering once again the duality between shift-symmetric scalars and Kalb-Ramond fields. Although in this work we have unveiled some of the properties of the dual formulation of the solid in terms of Kalb-Ramond fields, a more thorough analysis would be interesting to carry out, especially in terms of how the symmetries are realised in both formulations (for the solid and its more symmetric partners the fluids) and their potential relation to cosmological adiabatic modes.

After exploring the massless case at some length, we have extended our analysis to massive 2-form fields. The absence of an internal gauge symmetry prevents the construction of a non-trivially realised cosmological principle. The massive Kalb-Ramond field is now dual to a Proca field and we have briefly discussed the potential construction of cosmological solutions dual to known cosmologies supported by massive vector fields. We have however shown how the configurations supported by a homogeneous pure temporal vector field correspond to a singular dualisation in the simplest cases. From the massive 2-form field we have discussed how to appropriately take the massless limit in a way that preserves the number of propagating degrees of freedom. By this procedure, we have recovered the known result that the massless limit of a massive 2-form reduces to two decoupled sectors conformed by a massless vector field plus a shift symmetric scalar field. We have argued how the massive multi-Kalb-Ramond theories do permit non-trivial realisations of the cosmological principle and discussed how these configurations naturally lead to cosmologies with a second helicity-2 mode that permits having oscillations of gravitational waves. Finally, we have mentioned how to construct ghost-free non-minimal couplings. It is known that gravity theories formulated in geometrical frameworks beyond the Riemannian realm are plagued by ghost-like instabilities and the source of these pathologies can, in many cases, be traced back to a pathological character of a 2-form field, usually associated to the a certain sector of the torsion. In this sense, unveiling healthy non-minimal couplings for the 2-form can pave the way to constructing healthy gravity theories without the mentioned pathologies.

A natural question to ask is why should we care about these cosmological scenarios with 2-form fields when they can be dualised to seemingly simpler theories with scalar or vector fields. We can mention several reasons. Firstly, the dual formulations can unveil new properties that can be obscure with scalars or vector fields. One example is the case of the solid cosmology where the homogeneity requires a combination of spatial translations and internal translations in the target space, while the formulation with 2-forms makes the homogeneity trivial. The price to pay is having to work with the redundancies introduced by the gauge symmetries. Another reason to consider the dual formulation is the usual weak/strong coupling regimes of dual theories so the regime of strong coupling in the scalar or vector field formulations could be tackled by dualising the theories and working with weakly coupled 2-form fields. From a pure phenomenological standpoint alone, one can argue that theories that appear very natural in one formulation could be very contrived in the dual formulation so it is worth and desirable to explore the model-building potential in both descriptions. This is particularly important when considering matter couplings since the duality procedure typically generates non-local operators in the dual description from local interactions in the original formulation. Also regarding matter couplings, it is well-known that the kinetic self-interactions of the shift-symmetric scalar fields permit the presence of screening mechanisms of the $K-$mouflage type. This screening mechanism can of course be dualised and described in terms of 2-forms, but the resulting matter couplings will be non-local and an appropriate analysis of the screening becomes more contrived. On the other hand, 2-form fields naturally couple to strings so charged strings act as sources of the Kalb-Ramond field. In this situation, we will also have a $K-$mouflaged screening for the interaction between strings mediated by the Kalb-Ramond field, with potential phenomenological consequences for cosmic strings. The analogous screening in the dual description in terms of a scalar field would in this case be more contrived. We hope to return to these issues in a future work.

\acknowledgments
This work has been funded by Project PID2021-122938NB-I00 funded by the Spanish ``Ministerio de Ciencia e Innovaci\'on" and FEDER ``A way of making Europe". The work of K.A. was supported in part by Grants-in-Aid from the Scientific Research Fund of the Japan Society for the Promotion of Science, No.~20K14468. DF acknowledges support from the programme {\it Ayudas para Financiar la Contrataci\'on Predoctoral de Personal Investigador (ORDEN EDU/601/2020)} funded by Junta de Castilla y Le\'on and European Social Fund. 

\vspace{1cm}
\appendix
\noindent {\bf \Large{Appendix}}

\section{Duality in the presence of higher order derivatives}
\label{sec:AppHO}

In Sec. \ref{sec:weak/strong} we discussed how the dualisation in the presence of higher order terms cannot be established by following the same procedure as for the leading order terms with first order derivatives. In this appendix we will show more explicitly how the duality for higher order derivative theories of scalars and 2-forms fails and how one can find dual theories in those cases (which will thus describe different theories). Let us commence by considering the higher order Lagrangian for a scalar field described by the Lagrangian:
\be
\Lag=F(X,\Box\phi).
\label{eq:scalarHO}
\ee
We can reduce the order of the Lagrangian by introducing appropriate non-dynamical fields as follows:
\be
\Lag=F(X,\Sigma)+\lambda\big(\Box\phi-\Sigma\big).
\ee
If we integrate out the auxiliary field $\Sigma$ by solving its equation of motion so we have $\Sigma=\Sigma(X,\lambda)$, we obtain the equivalent Lagrangian:
\be
\Lag=-\partial_\mu\lambda\partial^\mu\phi+\mathcal{F}(X,\lambda)
\ee
with $\mathcal{F}=[F(X,\Sigma)-\lambda\Sigma]\vert_{\Sigma(X,\lambda)}$. This form of the Lagrangian makes apparent two important facts. Firstly, the absence of a proper kinetic term for $\lambda$, i.e., the Lagrangian does not contain $(\partial\lambda)^2$, prevents having a positive definite kinetic sector so, if both fields are dynamical, one of them must be a ghost. Secondly, the field $\lambda$ does not have a shift symmetry and this fact obstructs the dualisation procedure considered in this work. Notice that this was not obvious in the original form of the Lagrangian \eqref{eq:scalarHO} where there is a shift symmetry.

We can apply the general procedure to the following specific Lagrangian
\be
\Lag=-\frac12 (\partial \phi)^2+\frac{1}{2\Lambda^2}(\Box\phi)^2.
\ee
to obtain the equivalent Lagrangian
\be
\Lag=-\frac12 (\partial \varphi)^2+\frac12(\partial\lambda)^2-\frac12 \Lambda^2 \lambda^2
\ee
where we have diagonalised the kinetic term. We see explicitly how the theory propagates a ghost whose mass is precisely the scale associated with the higher order derivative term. 

We can perform a similar analysis for the higher order Lagrangian for a 2-form
\be
\Lag=G(H^2,\partial_\mu H^{\mu\alpha\beta}).
\ee
By introducing non-dynamical fields, we can rewrite the Lagrangian as
\be
\Lag=G(H^2,\Sigma^{\mu\nu})+\lambda_{\alpha\beta}\Big(\partial_\mu H^{\mu\alpha\beta}-\Sigma^{\alpha\beta}\Big).
\ee
Notice that both $\lambda_{\mu\nu}$ and $\Sigma_{\mu\nu}$ are antisymmetric fields. By integrating out $\Sigma_{\mu\nu}$ we obtain the equivalent Lagrangian
\be
\Lag=-\partial_{[\mu}\lambda_{\nu\rho]}H^{\mu\nu\rho}+\mathcal{G}(H^2,\lambda_{\mu\nu})
\ee
with 
\be
\mathcal{G}=\Big[G(H^2,\Sigma)-\lambda^{\mu\nu}\Sigma_{\mu\nu}\Big]_{\Sigma(H^2,\lambda)}.
\ee
Again, we see that the 2-form $\lambda_{\mu\nu}$ does not have a proper kinetic term so that the theory necessarily propagates a ghost because the kinetic matrix is not positive definite. Furthermore, $\lambda_{\mu\nu}$ does not have a gauge symmetry so the theory propagates four dof's as it corresponds to a massive plus a massless 2-forms. 

We can again apply the general procedure to the specific Lagrangian\footnote{The term $\partial_\mu H_{\nu\rho\sigma} \partial^\mu H^{\nu\rho\sigma}$ is equivalent to $(\partial_\mu H^{\mu\alpha\beta})^2$ via integration by parts so, at this order, this is the most general Lagrangian.}
\be
\Lag=-\frac{1}{12}H^2-\frac{1}{2\Lambda^2}(\partial_\mu H^{\mu\alpha\beta})^2
\ee
that admits the equivalent representation
\be
\Lag=-\frac{1}{12}H^2(B)+\frac13H^2(\lambda)-\frac12\Lambda^2\lambda_{\alpha\beta} \lambda^{\alpha\beta},
\ee
where we have diagonalised to exhibit the ghostly nature of one of the 2-form fields in an explicit manner and its mass $\Lambda^2$ that coincides with the scale of the higher order term. Being one of the 2-forms massive, we thus arrive at the same conclusion that the duality procedure to a higher derivative scalar field cannot be performed. At best, we could dualise the massless 2-form to a scalar and the massive 2-form to a massive vector field.

This simple example makes it apparent how the higher order derivative terms for the scalar field and the 2-forms introduce different numbers of dof's, thus obstructing the duality procedure. Of course, this result can be extended straightforwardly to general self-interacting fields.

It is important to notice that we can still obtain some sort of dual theories for both the scalar field and the 2-form theories with higher order derivatives. They just do not happen to be related in a simple manner in general. Let us corroborate this by dualising both theories. 

We will now consider the more general theory featuring second derivatives of a scalar field described by the following Lagrangian
\be
\Lag =F(\partial_\mu\phi,\partial_\alpha\partial_\beta\phi).
\ee
We first rewrite this Lagrangian in the first order form
\be
\Lag=\pi^{\mu}\Big(\partial_\mu\phi-A_\mu\Big)+\Sigma^{\mu\nu}\partial_{(\mu}A_{\nu)}+\mathcal{H}(A,\Sigma).
\ee
It is immediate to check that we can solve the non-dynamical equations for $\Sigma_{\mu\nu}$ and $\pi$ to obtain $A_\mu=\partial_\mu\phi$ and $\Sigma_{\mu\nu}=\Sigma_{\mu\nu}(\partial_\mu\phi,\partial_\alpha\partial_\beta\phi)$ that we can plug back to recover the original Lagrangian with $F=\Sigma^{\mu\nu}\partial_{(\mu}A_{\nu)}+\mathcal{H}(A,\Sigma)$. Alternatively, we can solve the equation for $\phi$ 
\be
\partial_\mu\pi^\mu=0\quad\Rightarrow\quad\pi^\mu=\epsilon^{\mu\nu\rho\sigma}\partial_{[\nu}\mathcal{B}_{\rho\sigma]}
\ee
for some 2-form field $\mathcal{B}_{\mu\nu}$. The Lagrangian can then be recast as
\be
\Lag=-\epsilon^{\mu\nu\rho\sigma}A_\mu\partial_{[\nu}\mathcal{B}_{\rho\sigma]}+\Sigma^{\mu\nu}\partial_{(\mu}A_{\nu)}+\mathcal{H}(A,\Sigma),
\label{eq:dualHOscalar}
\ee
where $\mathcal{B}$ represents the dual of $\phi$ regarding the first order operators. From here we can follow different paths to obtain equivalent {\it dual} theories depending on which of the fields we decide to integrate out, i.e., how we want to describe the higher order operators. Alternatively, we can use the Hodge dual $A_{\nu\rho\sigma}=\epsilon_{\mu\nu\rho\sigma}A^\mu$ to rewrite the Lagrangian as
\be
\Lag=-A^{\nu\rho\sigma}\partial_{[\nu}\mathcal{B}_{\rho\sigma]}-\frac16\partial_\mu\tilde{\Sigma}^{\mu\nu\rho\sigma}A_{\nu\rho\sigma}+\mathcal{H}(A,\tilde{\Sigma}),
\ee
where we have defined $\tilde{\Sigma}^{\mu}{}_{\rho\sigma\lambda}\equiv\Sigma^{\mu\nu}\epsilon_{\nu\rho\sigma\lambda}$. Again, we can integrate out different fields to reach different dualisations of the original higher order Lagrangian. 

We will not delve into the different options that can be explored, since it is not our purpose to perform a detailed analysis of the dualisations of higher order operators in this Appendix. Instead, we will now show how the analogous construction can be carried out for a 2-form. Thus, let us consider the Lagrangian
\be
\Lag= G(H_{\mu\nu\rho},\partial_{\alpha}H_{\mu\nu\rho}).
\ee
Its first order form can be written as
\be
\Lag=\Pi^{\mu\nu\rho}\Big(H_{\mu\nu\rho}-\mathcal{A}_{\mu\nu\rho}\Big)+\Sigma^{\mu\nu\rho\sigma}\partial_{\mu}\mathcal{A}_{\nu\rho\sigma}+\mathcal{H}(A,\Sigma),
\ee
from which we can recover the original higher order form by integrating out $\Pi^{\mu\nu\rho}$ and $\Sigma^{\mu\nu\rho\sigma}$. On the other hand, we can solve the equation for the 2-form $\partial_\mu\Pi^{\mu\nu\rho}=0$ that implies $\Pi^{\mu\nu\rho}=\epsilon^{\mu\nu\rho\lambda}\partial_\lambda\varphi$
to obtain the equivalent Lagrangian
\be
\Lag=-\epsilon^{\mu\nu\rho\lambda}\mathcal{A}_{\mu\nu\rho}\partial_\lambda\varphi+\Sigma^{\mu\nu\rho\sigma}\partial_{\mu}\mathcal{A}_{\nu\rho\sigma}+\mathcal{H}(\mathcal{A},\Sigma),
\ee
that can also be written in terms of $\mathcal{A}^\lambda=\frac16\epsilon^{\mu\nu\rho\lambda}\mathcal{A}_{\mu\nu\rho}$ as
\be
\Lag=-6\mathcal{A}^\lambda\partial_\lambda\varphi+\tilde{\Sigma}^{\mu\lambda}\partial_{\mu}\mathcal{A}_\lambda+\mathcal{H}(\mathcal{A},\tilde{\Sigma}),
\label{eq:dualHO2form}
\ee
with $\tilde{\Sigma}^\mu{}_\lambda\equiv\Sigma^{\mu\nu\rho\sigma}\epsilon_{\nu\rho\sigma\lambda}$. Again, the scalar field $\varphi$ represents the dual of the original 2-form, while the dualisation corresponding to the higher order operators can be performed in different ways depending on which non-dynamical field we decide to integrate out. For our purpose here, it is sufficient to notice that \eqref{eq:dualHOscalar} and \eqref{eq:dualHO2form} describe different theories and do not represent dual descriptions of the same theory as it was made manifest from our analysis above that showed the mismatch in the number of propagating dof's. We can insist once more that this is not really relevant from an EFT perspective. As a matter of fact, if we integrate out all the fields but the scalar and the 2-form (i.e., $A$ and $\Sigma$), it is easy to see that both theories share the same leading-order EFT below the corresponding ghost scale and the different dof's only manifest themselves in the higher order corrections. It would be interesting in any case to perform a more detailed analysis of the dual theories, expressed in e.g. \eqref{eq:dualHOscalar} and \eqref{eq:dualHO2form}, that we have obtained.

\section{On incomplete gauge fixing}
\label{sec:gaugefix}
In this appendix we will discuss how a part of the equations of motion is lost when we incompletely fix a gauge at the level of the action and how this missing equation of motion can be recovered. For simplicity, we consider the self-interacting Kalb-Ramond field in flat spacetime:
\be
\mS[B_{\mu\nu}] = \int \dd^4x F(Y)\,.
\ee
The general variation keeping the boundary term is given by
\be
\delta \mS = \int \dd^4 x \left[\frac{1}{2}\mathcal{E}^{\mu\nu}\delta B_{\mu\nu} -\frac{1}{2}\partial_{\mu}(F_Y H^{\mu\nu\rho} \delta B_{\nu\rho} ) \right]
\ee
where
\be
\mathcal{E}^{\mu\nu} \equiv \partial_{\rho}(F_Y H^{\mu\nu\rho} )
\,.
\ee
Hence, the Euler-Lagrange equation is $\mathcal{E}^{\mu\nu}=0$ which can be decomposed as 
\be
\mathcal{E}^{0i}=0 \,, \quad \mathcal{E}^{ij}=0
\,.
\ee
On the other hand, the variation must identically vanish for $\delta B_{\mu\nu}=2\partial_{[\mu}\theta_{\nu]}$ for any $\theta_{\mu}(t,\vec{x})$ by virtue of the gauge invariancee of the Lagrangian. We consider $\theta_{\mu}(t,\vec{x})$ with a compact support in spacetime to ignore the boundary term. Then, we find
\be
0\equiv \delta_{\theta} \mS = \int \dd^4 x \mathcal{E}^{\mu\nu}\partial_{\mu} \theta_{\nu}
=-\int \dd^4 x \partial_{\mu}\mathcal{E}^{\mu\nu} \theta_{\nu}
\,,
\ee
yielding the identities
\bea
\partial_{\mu}\mathcal{E}^{\mu 0} &=& \partial_i \mathcal{E}^{i0}  \equiv 0
\,, \label{2formidentity1}
\\
\partial_{\mu}\mathcal{E}^{\mu i} &=& \partial_0 \mathcal{E}^{0i} + \partial_i \mathcal{E}^{ij} \equiv 0
\,.
\label{2formidentity2}
\eea
The first identity \eqref{2formidentity1} shows that the longitudinal part of $\mathcal{E}^{i0}$ is a redundant equation and \eqref{2formidentity2} concludes that the transverse part of $\mathcal{E}^{i0}$ is related to $\mathcal{E}^{ij}$.

Next, we consider the gauge-fixed action
\be
\mS_{\rm gf}[B_{ij}] = \int \dd^4x F(Y)|_{B_{0i}=0}
\label{gf2form}
\ee
whose variation leads to
\be
\delta \mS_{\rm gf} = \int \dd^4 x \left[\frac{1}{2}\mathcal{E}^{ij} \delta B_{ij} -\frac{1}{2}\partial_{\mu}(F_Y H^{\mu ij} \delta B_{ij} ) \right]_{B_{0i}=0 }
\,.
\ee
Therefore, we only find the Euler-Lagrange equation 
\be
\mathcal{E}^{ij}|_{B_{0i}=0} = 0
\ee
from the gauge-fixed action while the original gauge-fixed equations are
\be
\mathcal{E}^{0i}|_{B_{0i}=0}=0 \,, \quad \mathcal{E}^{ij}|_{B_{0i}=0}=0
\,.
\ee
The equation $\mathcal{E}^{0i}|_{B_{0i}=0}=0$ is lost in the gauge-fixed action \eqref{gf2form}. One may notice that the identities \eqref{2formidentity1} and \eqref{2formidentity2} may be used to recover the missing equations of motion. Indeed, we know that the longitudinal part of $\mathcal{E}^{0i}|_{B_{0i}=0}=0$ is a redundant equation, so we only need to recover the transverse part of $\mathcal{E}^{0i}|_{B_{0i}=0}=0$ which we denote by $\vec{\mathcal{E}}^T|_{B_{0i}=0}=0$. By using \eqref{2formidentity2}, we find
\be
\mathcal{E}^{ij}|_{B_{0i}=0} = 0 \implies \partial_0 \vec{\mathcal{E}}^T|_{B_{0i}=0} = 0
\,.
\ee
However, this is a differential equation of $\mathcal{E}^{0i}|_{B_{0i}=0}$ and the generic solution is
\be
\vec{\mathcal{E}}^T|_{B_{0i}=0} = \vec{\mathcal{J}}^T (\vec{x})\,, \quad {\rm s.t.} \quad \nabla \cdot \vec{\mathcal{J}}^T(\vec{x})=0
\,,
\ee
where $\vec{\mathcal{J}}^T(\vec{x})$ is an integration constant. Hence, the gauge-fixed action is consistent with the original action if and only if we set $\vec{\mathcal{J}}^T(\vec{x})=0$.

The apparent freedom in the choice of $\vec{\mathcal{J}}^T$ is related to the undetermined part of the gauge parameter. The gauge-fixed action \eqref{gf2form} enjoys the (infinite number of) global symmetries
\be
B_{ij}(t,\vec{x}) \to B_{ij}(t,\vec{x}) + 2\partial_{[i}\theta^T_{j]}(\vec{x})\,, \quad {\rm s.t.} \quad \partial^i \theta^{T}_i(\vec{x}) =0
\,,
\label{2formglobal}
\ee
which is originally a part of the gauge symmetry of the 2-form; thus, the gauge condition $B_{0i}=0$ is not a complete gauge-fixing. By substituting $\delta B_{ij}=2\partial_{[i}\theta^T_{j]}(\vec{x})$ with a compact support $\theta^T_i(\vec{x})$ in space, we obtain the identity
\bea
0\equiv \delta_{\theta^T} \mS_{\rm gf} &= & \int \dd^4 x \left[ \mathcal{E}^{ij}\partial_i\theta^T_j - \partial_0 (F_Y H^{0ij}) \partial_i \theta^T_j \right]_{B_{0i}=0}
\nn
&=& \int \dd^4 x \theta^T_j(\vec{x}) \left[ -\partial_i \mathcal{E}^{ij}  - \partial_0 \mathcal{J}^{Ti} \right]_{B_{0i}=0} 
\label{gf2formid}
\eea
where 
\be
\mathcal{J}^{Ti} \equiv \partial_j(F_Y H^{0ij})|_{B_{0i}=0} 
\,,
\ee
is interpreted as a conserved charge associated with the global symmetry \eqref{2formglobal}. If we only know the gauge-fixed action \eqref{2formglobal}, the symmetry of the theory is the global one and we may not exclude the existence of the charge. However, the transformation \eqref{2formglobal} should be regarded as a part of the original gauge transformation, leading to a consistency condition: we have to restrict solutions to ones having no charge $\mathcal{J}^{Ti} = 0$ at an initial spacelike hypersurface. Indeed, the condition $\mathcal{J}^{Ti} = 0$ coincides with the missing equation of motion $\vec{\mathcal{E}}^T|_{B_{0i}=0} =0 $ and the identity \eqref{gf2formid} guarantees that $\mathcal{J}^{Ti} = 0$ holds at any time when it is satisfied at the initial hypersurface. This concludes that the gauge-fixed action \eqref{gf2form} can reproduce the consistent equations of motion by choosing the appropriate initial condition.

\bibliography{bibdualities}

\providecommand{\href}[2]{#2}\begingroup\raggedright\begin{thebibliography}{10}

\bibitem{Secrest:2022uvx}
N.~J. Secrest, S.~von Hausegger, M.~Rameez, R.~Mohayaee, and S.~Sarkar, {\it {A
  Challenge to the Standard Cosmological Model}},  {\em Astrophys. J. Lett.}
  {\bf 937} (2022), no.~2 L31, [\href{http://arxiv.org/abs/2206.05624}{{\tt
  arXiv:2206.05624}}].

\bibitem{Dalang:2021ruy}
C.~Dalang and C.~Bonvin, {\it {On the kinematic cosmic dipole tension}},  {\em
  Mon. Not. Roy. Astron. Soc.} {\bf 512} (2022), no.~3 3895--3905,
  [\href{http://arxiv.org/abs/2111.03616}{{\tt arXiv:2111.03616}}].

\bibitem{Guandalin:2022tyl}
C.~Guandalin, J.~Piat, C.~Clarkson, and R.~Maartens, {\it {Theoretical
  systematics in testing the Cosmological Principle with the kinematic quasar
  dipole}},  \href{http://arxiv.org/abs/2212.04925}{{\tt arXiv:2212.04925}}.

\bibitem{Sorrenti:2022zat}
F.~Sorrenti, R.~Durrer, and M.~Kunz, {\it {The Dipole of the Pantheon+SH0ES
  Data}},  \href{http://arxiv.org/abs/2212.10328}{{\tt arXiv:2212.10328}}.

\bibitem{Atrio-Barandela:2014nda}
F.~Atrio-Barandela, A.~Kashlinsky, H.~Ebeling, D.~J. Fixsen, and D.~Kocevski,
  {\it {Probing the Dark Flow Signal in Wmap 9 -year and Planck Cosmic
  Microwave Background Maps}},  {\em Astrophys. J.} {\bf 810} (2015), no.~2
  143, [\href{http://arxiv.org/abs/1411.4180}{{\tt arXiv:1411.4180}}].

\bibitem{Cervero:1978db}
J.~Cervero and L.~Jacobs, {\it {Classical {Yang-Mills} Fields in a
  Robertson-walker Universe}},  {\em Phys. Lett. B} {\bf 78} (1978) 427--429.

\bibitem{Galtsov:1991un}
D.~V. Galtsov and M.~S. Volkov, {\it {Yang-Mills cosmology: Cold matter for a
  hot universe}},  {\em Phys. Lett. B} {\bf 256} (1991) 17--21.

\bibitem{Darian:1996mb}
B.~K. Darian and H.~P. Kunzle, {\it {Cosmological Einstein Yang-Mills
  equations}},  {\em J. Math. Phys.} {\bf 38} (1997) 4696--4713,
  [\href{http://arxiv.org/abs/gr-qc/9610026}{{\tt gr-qc/9610026}}].

\bibitem{Maleknejad:2011jw}
A.~Maleknejad and M.~M. Sheikh-Jabbari, {\it {Gauge-flation: Inflation From
  Non-Abelian Gauge Fields}},  {\em Phys. Lett. B} {\bf 723} (2013) 224--228,
  [\href{http://arxiv.org/abs/1102.1513}{{\tt arXiv:1102.1513}}].

\bibitem{Maleknejad:2012fw}
A.~Maleknejad, M.~M. Sheikh-Jabbari, and J.~Soda, {\it {Gauge Fields and
  Inflation}},  {\em Phys. Rept.} {\bf 528} (2013) 161--261,
  [\href{http://arxiv.org/abs/1212.2921}{{\tt arXiv:1212.2921}}].

\bibitem{Armendariz-Picon:2004say}
C.~Armendariz-Picon, {\it {Could dark energy be vector-like?}},  {\em JCAP}
  {\bf 07} (2004) 007, [\href{http://arxiv.org/abs/astro-ph/0405267}{{\tt
  astro-ph/0405267}}].

\bibitem{Endlich:2012pz}
S.~Endlich, A.~Nicolis, and J.~Wang, {\it {Solid Inflation}},  {\em JCAP} {\bf
  10} (2013) 011, [\href{http://arxiv.org/abs/1210.0569}{{\tt
  arXiv:1210.0569}}].

\bibitem{Bucher:1998mh}
M.~Bucher and D.~N. Spergel, {\it {Is the dark matter a solid?}},  {\em Phys.
  Rev. D} {\bf 60} (1999) 043505,
  [\href{http://arxiv.org/abs/astro-ph/9812022}{{\tt astro-ph/9812022}}].

\bibitem{Gruzinov:2004ty}
A.~Gruzinov, {\it {Elastic inflation}},  {\em Phys. Rev. D} {\bf 70} (2004)
  063518, [\href{http://arxiv.org/abs/astro-ph/0404548}{{\tt
  astro-ph/0404548}}].

\bibitem{Piazza:2017bsd}
F.~Piazza, D.~Pirtskhalava, R.~Rattazzi, and O.~Simon, {\it {Gaugid
  inflation}},  {\em JCAP} {\bf 11} (2017) 041,
  [\href{http://arxiv.org/abs/1706.03402}{{\tt arXiv:1706.03402}}].

\bibitem{Nicolis:2015sra}
A.~Nicolis, R.~Penco, F.~Piazza, and R.~Rattazzi, {\it {Zoology of condensed
  matter: Framids, ordinary stuff, extra-ordinary stuff}},  {\em JHEP} {\bf 06}
  (2015) 155, [\href{http://arxiv.org/abs/1501.03845}{{\tt arXiv:1501.03845}}].

\bibitem{Kang:2015uha}
J.~Kang and A.~Nicolis, {\it {Platonic solids back in the sky: Icosahedral
  inflation}},  {\em JCAP} {\bf 03} (2016) 050,
  [\href{http://arxiv.org/abs/1509.02942}{{\tt arXiv:1509.02942}}].

\bibitem{Cremmer:1973mg}
E.~Cremmer and J.~Scherk, {\it {Spontaneous dynamical breaking of gauge
  symmetry in dual models}},  {\em Nucl. Phys. B} {\bf 72} (1974) 117--124.

\bibitem{Stein-Schabes:1986owe}
J.~A. Stein-Schabes and M.~Gleiser, {\it {Einstein-Kalb-Ramond Cosmology}},
  {\em Phys. Rev. D} {\bf 34} (1986) 3242.

\bibitem{Copeland:1994km}
E.~J. Copeland, A.~Lahiri, and D.~Wands, {\it {String cosmology with a time
  dependent antisymmetric tensor potential}},  {\em Phys. Rev. D} {\bf 51}
  (1995) 1569--1576, [\href{http://arxiv.org/abs/hep-th/9410136}{{\tt
  hep-th/9410136}}].

\bibitem{Elizalde:2018rmz}
E.~Elizalde, S.~D. Odintsov, T.~Paul, and D.~S\'aez-Chill\'on~G\'omez, {\it
  {Inflationary universe in $F(R)$ gravity with antisymmetric tensor fields and
  their suppression during its evolution}},  {\em Phys. Rev. D} {\bf 99}
  (2019), no.~6 063506, [\href{http://arxiv.org/abs/1811.02960}{{\tt
  arXiv:1811.02960}}].

\bibitem{Curtright:1980yj}
T.~L. Curtright and P.~G.~O. Freund, {\it {Massive dual fields}},  {\em Nucl.
  Phys. B} {\bf 172} (1980) 413--424.

\bibitem{Matsuo:2021xas}
Y.~Matsuo and A.~Sugamoto, {\it {Note on a description of a perfect fluid by
  the Kalb\textendash{}Ramond field}},  {\em PTEP} {\bf 2021} (2021), no.~12
  12C104.

\bibitem{Horn:2015zna}
B.~Horn, A.~Nicolis, and R.~Penco, {\it {Effective string theory for vortex
  lines in fluids and superfluids}},  {\em JHEP} {\bf 10} (2015) 153,
  [\href{http://arxiv.org/abs/1507.05635}{{\tt arXiv:1507.05635}}].

\bibitem{Arkani-Hamed:2003pdi}
N.~Arkani-Hamed, H.-C. Cheng, M.~A. Luty, and S.~Mukohyama, {\it {Ghost
  condensation and a consistent infrared modification of gravity}},  {\em JHEP}
  {\bf 05} (2004) 074, [\href{http://arxiv.org/abs/hep-th/0312099}{{\tt
  hep-th/0312099}}].

\bibitem{Arkani-Hamed:2003juy}
N.~Arkani-Hamed, P.~Creminelli, S.~Mukohyama, and M.~Zaldarriaga, {\it {Ghost
  inflation}},  {\em JCAP} {\bf 04} (2004) 001,
  [\href{http://arxiv.org/abs/hep-th/0312100}{{\tt hep-th/0312100}}].

\bibitem{Babichev:2016hys}
E.~Babichev, {\it {Formation of caustics in k-essence and Horndeski theory}},
  {\em JHEP} {\bf 04} (2016) 129, [\href{http://arxiv.org/abs/1602.00735}{{\tt
  arXiv:1602.00735}}].

\bibitem{Babichev:2017lrx}
E.~Babichev and S.~Ramazanov, {\it {Caustic free completion of pressureless
  perfect fluid and k-essence}},  {\em JHEP} {\bf 08} (2017) 040,
  [\href{http://arxiv.org/abs/1704.03367}{{\tt arXiv:1704.03367}}].

\bibitem{Babichev:2018twg}
E.~Babichev, S.~Ramazanov, and A.~Vikman, {\it {Recovering $P(X)$ from a
  canonical complex field}},  {\em JCAP} {\bf 11} (2018) 023,
  [\href{http://arxiv.org/abs/1807.10281}{{\tt arXiv:1807.10281}}].

\bibitem{Mizuno:2019pcm}
S.~Mizuno, S.~Mukohyama, S.~Pi, and Y.-L. Zhang, {\it {Hyperbolic field space
  and swampland conjecture for DBI scalar}},  {\em JCAP} {\bf 09} (2019) 072,
  [\href{http://arxiv.org/abs/1905.10950}{{\tt arXiv:1905.10950}}].

\bibitem{Mukohyama:2020lsu}
S.~Mukohyama and R.~Namba, {\it {Partial UV Completion of $P(X)$ from a Curved
  Field Space}},  {\em JCAP} {\bf 02} (2021) 001,
  [\href{http://arxiv.org/abs/2010.09184}{{\tt arXiv:2010.09184}}].

\bibitem{Aoki:2021ffc}
K.~Aoki, S.~Mukohyama, and R.~Namba, {\it {Positivity vs. Lorentz-violation: an
  explicit example}},  {\em JCAP} {\bf 10} (2021) 079,
  [\href{http://arxiv.org/abs/2107.01755}{{\tt arXiv:2107.01755}}].

\bibitem{Dubovsky:2005xd}
S.~Dubovsky, T.~Gregoire, A.~Nicolis, and R.~Rattazzi, {\it {Null energy
  condition and superluminal propagation}},  {\em JHEP} {\bf 03} (2006) 025,
  [\href{http://arxiv.org/abs/hep-th/0512260}{{\tt hep-th/0512260}}].

\bibitem{Carter:1994rv}
B.~Carter, {\it {Axionic vorticity variational formulation for relativistic
  perfect fluids}},  {\em Class. Quant. Grav.} {\bf 11} (1994) 2013--2030.

\bibitem{Carter:1995mj}
B.~Carter and D.~Langlois, {\it {Kalb-Ramond coupled vortex fibration model for
  relativistic superfluid dynamics}},  {\em Nucl. Phys. B} {\bf 454} (1995)
  402--424, [\href{http://arxiv.org/abs/hep-th/9611082}{{\tt hep-th/9611082}}].

\bibitem{Garcia-Saenz:2017wzf}
S.~Garcia-Saenz, E.~Mitsou, and A.~Nicolis, {\it {A multipole-expanded
  effective field theory for vortex ring-sound interactions}},  {\em JHEP} {\bf
  02} (2018) 022, [\href{http://arxiv.org/abs/1709.01927}{{\tt
  arXiv:1709.01927}}].

\bibitem{Pajer:2018egx}
E.~Pajer and D.~Stefanyszyn, {\it {Symmetric Superfluids}},  {\em JHEP} {\bf
  06} (2019) 008, [\href{http://arxiv.org/abs/1812.05133}{{\tt
  arXiv:1812.05133}}].

\bibitem{Afshordi:2006ad}
N.~Afshordi, D.~J.~H. Chung, and G.~Geshnizjani, {\it {Cuscuton: A Causal Field
  Theory with an Infinite Speed of Sound}},  {\em Phys. Rev. D} {\bf 75} (2007)
  083513, [\href{http://arxiv.org/abs/hep-th/0609150}{{\tt hep-th/0609150}}].

\bibitem{Gomes:2017tzd}
H.~Gomes and D.~C. Guariento, {\it {Hamiltonian analysis of the cuscuton}},
  {\em Phys. Rev. D} {\bf 95} (2017), no.~10 104049,
  [\href{http://arxiv.org/abs/1703.08226}{{\tt arXiv:1703.08226}}].

\bibitem{Born:1934gh}
M.~Born and L.~Infeld, {\it {Foundations of the new field theory}},  {\em Proc.
  Roy. Soc. Lond. A} {\bf 144} (1934), no.~852 425--451.

\bibitem{Cheung:2007st}
C.~Cheung, P.~Creminelli, A.~L. Fitzpatrick, J.~Kaplan, and L.~Senatore, {\it
  {The Effective Field Theory of Inflation}},  {\em JHEP} {\bf 03} (2008) 014,
  [\href{http://arxiv.org/abs/0709.0293}{{\tt arXiv:0709.0293}}].

\bibitem{Aoki:2021wew}
K.~Aoki, M.~A. Gorji, S.~Mukohyama, and K.~Takahashi, {\it {The effective field
  theory of vector-tensor theories}},  {\em JCAP} {\bf 01} (2022), no.~01 059,
  [\href{http://arxiv.org/abs/2111.08119}{{\tt arXiv:2111.08119}}].

\bibitem{Aoki:2022ipw}
K.~Aoki, M.~A. Gorji, S.~Mukohyama, and K.~Takahashi, {\it {Effective Field
  Theory of Gravitating Continuum: Solids, Fluids, and Aether Unified}},
  \href{http://arxiv.org/abs/2204.06672}{{\tt arXiv:2204.06672}}.

\bibitem{Motohashi:2016prk}
H.~Motohashi, T.~Suyama, and K.~Takahashi, {\it {Fundamental theorem on gauge
  fixing at the action level}},  {\em Phys. Rev. D} {\bf 94} (2016), no.~12
  124021, [\href{http://arxiv.org/abs/1608.00071}{{\tt arXiv:1608.00071}}].

\bibitem{Cembranos:2012kk}
J.~A.~R. Cembranos, C.~Hallabrin, A.~L. Maroto, and S.~J.~N. Jareno, {\it
  {Isotropy theorem for cosmological vector fields}},  {\em Phys. Rev. D} {\bf
  86} (2012) 021301, [\href{http://arxiv.org/abs/1203.6221}{{\tt
  arXiv:1203.6221}}].

\bibitem{Cembranos:2012ng}
J.~A.~R. Cembranos, A.~L. Maroto, and S.~J. N\'u\~nez Jare\~no, {\it {Isotropy
  theorem for cosmological Yang-Mills theories}},  {\em Phys. Rev. D} {\bf 87}
  (2013), no.~4 043523, [\href{http://arxiv.org/abs/1212.3201}{{\tt
  arXiv:1212.3201}}].

\bibitem{Cembranos:2013cba}
J.~A.~R. Cembranos, A.~L. Maroto, and S.~J. N\'u\~nez Jare\~no, {\it {Isotropy
  theorem for arbitrary-spin cosmological fields}},  {\em JCAP} {\bf 03} (2014)
  042, [\href{http://arxiv.org/abs/1311.1402}{{\tt arXiv:1311.1402}}].

\bibitem{BeltranJimenez:2019xxx}
J.~Beltr\'an~Jim\'enez, J.~M. Ezquiaga, and L.~Heisenberg, {\it {Probing
  cosmological fields with gravitational wave oscillations}},  {\em JCAP} {\bf
  04} (2020) 027, [\href{http://arxiv.org/abs/1912.06104}{{\tt
  arXiv:1912.06104}}].

\bibitem{Ezquiaga:2021ler}
J.~M. Ezquiaga, W.~Hu, M.~Lagos, and M.-X. Lin, {\it {Gravitational wave
  propagation beyond general relativity: waveform distortions and echoes}},
  {\em JCAP} {\bf 11} (2021), no.~11 048,
  [\href{http://arxiv.org/abs/2108.10872}{{\tt arXiv:2108.10872}}].

\bibitem{Caldwell:2016sut}
R.~R. Caldwell, C.~Devulder, and N.~A. Maksimova, {\it {Gravitational
  wave\textendash{}Gauge field oscillations}},  {\em Phys. Rev. D} {\bf 94}
  (2016), no.~6 063005, [\href{http://arxiv.org/abs/1604.08939}{{\tt
  arXiv:1604.08939}}].

\bibitem{BeltranJimenez:2018ymu}
J.~Beltr\'an~Jim\'enez and L.~Heisenberg, {\it {Non-trivial gravitational waves
  and structure formation phenomenology from dark energy}},  {\em JCAP} {\bf
  09} (2018) 035, [\href{http://arxiv.org/abs/1806.01753}{{\tt
  arXiv:1806.01753}}].

\bibitem{Caldwell:2018feo}
R.~R. Caldwell and C.~Devulder, {\it {Gravitational Wave Opacity from Gauge
  Field Dark Energy}},  {\em Phys. Rev. D} {\bf 100} (2019), no.~10 103510,
  [\href{http://arxiv.org/abs/1802.07371}{{\tt arXiv:1802.07371}}].

\bibitem{Horndeski:1976gi}
G.~W. Horndeski, {\it {Conservation of Charge and the Einstein-Maxwell Field
  Equations}},  {\em J. Math. Phys.} {\bf 17} (1976) 1980--1987.

\bibitem{Yoshida:2019dxu}
D.~Yoshida, {\it {2-form gauge theory dual to scalar-tensor theory}},  {\em
  Phys. Rev. D} {\bf 100} (2019) 084047,
  [\href{http://arxiv.org/abs/1906.02462}{{\tt arXiv:1906.02462}}].

\bibitem{Takahashi:2019vax}
K.~Takahashi and D.~Yoshida, {\it {Ghost-free resummation of gravitational
  interactions of a two-form gauge field}},  {\em Phys. Rev. D} {\bf 101}
  (2020), no.~2 024049, [\href{http://arxiv.org/abs/1910.12508}{{\tt
  arXiv:1910.12508}}].

\bibitem{Bettoni:2015wla}
D.~Bettoni and S.~Liberati, {\it {Dynamics of non-minimally coupled perfect
  fluids}},  {\em JCAP} {\bf 08} (2015) 023,
  [\href{http://arxiv.org/abs/1502.06613}{{\tt arXiv:1502.06613}}].

\end{thebibliography}\endgroup

\end{document}